\documentclass[11pt]{article}

\usepackage[latin1]{inputenc}\usepackage{epsfig}\usepackage{amsfonts}

\title{
  Gaussian-Perturbative Calculations \\
  with a \\
  Homogeneous External Source }

\author{
  \Large Jorge L. deLyra \\ Department of Mathematical Physics \\
  Physics Institute \\ University of São Paulo }

\date{March 17, 2014}

\newcommand{\ii}{\mbox{\boldmath$\imath$}}

\newcommand{\tvphi}{\widetilde\varphi}

\newcommand{\OO}{{\mathbf O}}

\newcommand{\sN}{\mathfrak{N}}

\begin{document}\maketitle

\begin{abstract}
  \noindent
  We derive the equation of the critical curve and calculate the
  renormalized masses of the $SO(\sN)$-symmetric $\lambda\phi^{4}$ model
  in the presence of a homogeneous external source. We do this using the
  Gaussian-Perturbative approximation on finite lattices and explicitly
  taking the continuum limit. No disabling divergences are found in the
  final results, and no renormalization is necessary. We show that the
  results give a complete description of the critical behavior of the
  model and of the phenomenon of spontaneous symmetry breaking, at the
  quantum-field-theoretical level.
  
  We show that the renormalized masses depend on the external source, and
  point out the consequences of that fact for the design of computer
  simulations of the model. We point out a simple but interesting
  consequence of the results, regarding the role of the $\lambda\phi^{4}$
  model in the Standard Model of high-energy particle physics. Using the
  experimentally known values of the mass and of the expectation value of
  the Higgs field, we determine uniquely the values of the {\em bare}
  dimensionless parameters $\alpha$ and $\lambda$ of the model, which turn
  out to be small numbers, significantly less that one.
  
\end{abstract}

\section{Introduction}

Years ago we introduced a calculational technique that was quite
successful in describing the critical behavior of the $SO(\sN)$-symmetric
Euclidean $\lambda\phi^{4}$ model in $d\geq 3$ spacetime
dimensions~\cite{pertheory}. Some quantities were calculated in $d=4$ and
compared to the results of computer simulations, yielding surprisingly
good results, and describing reliably the most important qualitative
aspects of the model. The observables calculated where the expectation
value of the field, which is the order parameter of the critical
transition of the model and describes the phenomenon of spontaneous
symmetry breaking, and the two-point function, from which one can get the
renormalized masses and hence the correlation lengths, in both phases of
the model.

Although inspired by and superficially similar to perturbation theory, the
technique can handle a phenomenon such as spontaneous symmetry breaking,
which is usually considered to be out of reach for plain perturbation
theory. The innovative and essential aspect of the technique is the use of
certain self-consistency conditions within a framework similar to that of
perturbation theory. The technique would be better described as a Gaussian
approximation rather than a perturbative expansion. As such, it is able to
produce good predictions for the one-point and two-point observables,
since these are the moments present in the Gaussian distribution, but
should not be expected to go much further than that. For lack of a better
name, we shall refer to it as the Gaussian-Perturbative approximation.

The important role that the four-component $\lambda\phi^{4}$ model plays
in the Standard Model of high-energy particle physics makes it certainly
interesting to learn more about it. In this paper we extend the
Gaussian-Perturbative technique introduced in~\cite{pertheory} to the same
model in the presence of external sources. These external sources are not
thought of merely as analytical devices used to extract the Green's
functions from the functional generators of the model, and to be put to
zero afterwards. Instead, they are thought of as actual physical sources
of particles in the model. One important objective is to determine how the
introduction of the external sources affects the values of the
renormalized masses in either phase of the model.

These are analytical calculations performed on the Euclidean lattice,
which therefore allow us to discuss, and to explicitly take, specific
continuum limits in the quantum theory. As we will see, there is no need
for perturbative renormalization, or for any regulation mechanism other
than the lattice where the model is defined. All calculations on finite
lattices are ordinary straightforward manipulations. Although there are
some quantities that do diverge in the continuum limit, they all cancel
off from the observables before the limit is taken.

\section{The Model}\label{TheModel}

Let us start by giving the definition of the model, in the classical and
quantum domains, and then quickly reviewing the Gaussian-Perturbative
approximation. Consider then the Euclidean quantum field theories of an
$SO(\sN)$-symmetric set of scalar fields $\vec{\phi}(x_{\mu})$ defined
within a periodical cubic box of side $L$ in $d$ dimensions by the
classical action

\noindent
\begin{eqnarray*}
  S\!\left[\raisebox{-0.3ex}{$\vec{\phi}$}\right]
  & = &
  \oint_{L^{d}}{\rm d}^{d}x
  \left\{
    \rule{0em}{4ex}
    \frac{1}{2}
    \sum_{\nu}^{d}
    \left[
      \partial_{\nu}\vec{\phi}(x_{\mu})
      \cdot
      \partial_{\nu}\vec{\phi}(x_{\mu})
    \right]
    +
    \frac{m^{2}}{2}\,
    \left[\vec{\phi}(x_{\mu})\cdot\vec{\phi}(x_{\mu})\right]
    +
  \right.
  \\
  &   &
  \hspace{4.0em}
  \left.
    \rule{0em}{4ex}
    +
    \frac{\Lambda}{4}
    \left[\vec{\phi}(x_{\mu})\cdot\vec{\phi}(x_{\mu})\right]^{2}
    -
    J_{0}\phi_{\sN}(x_{\mu})
  \right\},
\end{eqnarray*}

\noindent
where $d\geq 3$. This is the usual form of the $SO(\sN)$-symmetric
$\lambda\phi^{4}$ model in the classical continuum, with an external
source $J_{0}$, which by assumption is a constant. The vector notation
$\vec{\phi}(x_{\mu})$ is shorthand for

\begin{displaymath}
  \vec{\phi}(x_{\mu})
  =
  \left(
    \phi_{1}(x_{\mu}),
    \phi_{2}(x_{\mu}),
    \ldots,
    \phi_{\sN}(x_{\mu})
  \right),
\end{displaymath}

\noindent
and the dot-product notation represents the scalar product of vectors in
the internal $SO(\sN)$ space, that is a sum over $i=1,2,\ldots,\sN$,

\begin{displaymath}
  \vec{\phi}(x_{\mu})\cdot\vec{\phi}(x_{\mu})
  =
  \sum_{i}^{\sN}
  \left[\phi_{i}(x_{\mu})\right]^{2}.
\end{displaymath}

\noindent
In this action the quantity $J_{0}$ is a homogeneous external source
associated with the $\phi_{\sN}(x_{\mu})$ field component. Its
introduction breaks the $SO(\sN)$ symmetry, of course, and causes the
generation of a non-zero expectation value for the $\phi_{\sN}(x_{\mu})$
field component.

In order to use the definition of the quantum theory on a cubical lattice
of size $L$ with $N$ sites along each direction, with lattice spacing
$a=L/N$, we consider the corresponding lattice action

\noindent
\begin{eqnarray*}
  S_{N}[\vec{\varphi}]
  & = &
  \sum_{n_{\mu}}^{N^{d}}
  \left\{
    \rule{0em}{4ex}
    \frac{1}{2}
    \sum_{\nu}^{d}
    \left[
      \Delta_{\nu}\vec{\varphi}(n_{\mu})
      \cdot
      \Delta_{\nu}\vec{\varphi}(n_{\mu})
    \right]
    +
    \frac{\alpha}{2}
    \left[\vec{\varphi}(n_{\mu})\cdot\vec{\varphi}(n_{\mu})\right]
    +
  \right.
  \\
  &   &
  \hspace{2.0em}
  \left.
    \rule{0em}{4ex}
    +
    \frac{\lambda}{4}
    \left[\vec{\varphi}(n_{\mu})\cdot\vec{\varphi}(n_{\mu})\right]^{2}
    -
    j_{0}\varphi_{\sN}(n_{\mu})
  \right\},
\end{eqnarray*}

\noindent
where all quantities are now dimensionless, defined by the appropriate
scalings,

\noindent
\begin{eqnarray}\label{Scalings}
  \varphi_{i}(n_{\mu})
  & = &
  a^{(d-2)/2}\phi_{i}(x_{\mu}),
  \nonumber\\
  n_{\mu}
  & = &
  a^{-1}x_{\mu},
  \nonumber\\
  \alpha
  & = &
  a^{2}m^{2},
  \\
  \lambda
  & = &
  a^{4-d}\Lambda,
  \nonumber\\
  j_{0}
  & = &
  a^{(d+2)/2}J_{0}.
  \nonumber
\end{eqnarray}

\noindent
In order for the model to be stable we must have $\lambda\geq 0$ and, in
addition to this, if $\lambda=0$ then we must also have $\alpha\geq 0$. Up
to this point there are no further constraints on the real parameters
$\alpha$ and $\lambda$.

When possible, the summations are notated, in the subscript, by the
variable which is being summed over, and, in the superscript, by the
number of terms in the sum. The integer coordinates $n_{\mu}$ are taken to
vary as symmetrically as possible around the origin $n_{\mu}=0_{\mu}$,
that is we have $n_{\mu}=n_{\rm min},\ldots,0,\ldots,n_{\rm max}$ with
certain values of $n_{\rm min}$ and $n_{\rm max}$ that depend on the
parity of $N$,

\begin{displaymath}
  n_{\mu}
  =
  -\,\frac{N-1}{2},\ldots,0,\ldots,\frac{N-1}{2},
\end{displaymath}

\noindent
for odd $N$, and

\begin{displaymath}
  n_{\mu}
  =
  -\,\frac{N}{2}+1,\ldots,0,\ldots,\frac{N}{2},
\end{displaymath}

\noindent
for even $N$, in either case for all values of $\mu=1,\ldots,d$.

In this paper we will perform the calculations of the critical line and of
the renormalized masses in a situation in which we have, in terms of the
dimensionfull field $\vec{\phi}(x_{\mu})$, for $i=1,\ldots,\sN-1$,

\begin{displaymath}
  \left\langle\phi_{i}(x_{\mu})\right\rangle
  =
  0,
\end{displaymath}

\noindent
and, for $i=\sN$,

\begin{displaymath}
  \left\langle\phi_{\sN}(x_{\mu})\right\rangle
  =
  V_{0},
\end{displaymath}

\noindent
where $V_{0}$ is a constant with the physical dimensions of the field
$\phi_{\sN}(x_{\mu})$. In terms of the dimensionless field
$\vec{\varphi}(n_{\mu})$ we have for the only non-trivial condition

\begin{displaymath}
  \left\langle\varphi_{\sN}(n_{\mu})\right\rangle
  =
  v_{0},
\end{displaymath}

\noindent
where the dimensionless constant is given by $v_{0}=a^{(d-2)/2}V_{0}$.

We will consider continuum limits in which we have both $N\to\infty$ and
$L\to\infty$. In order to do this we will choose to make $L$ increase as
$\sqrt{N}$ and $a$ decrease as $\sqrt{N}$, so that we still have
$a=L/N$. The calculations on finite lattices will be performed with
periodical boundary conditions, with the understanding that at the end of
the day such a limit is to be taken.

Observe that we are specifying the value of the expectation value $v_{0}$
of $\varphi_{\sN}(n_{\mu})$ rather than the value of the corresponding
external source $j_{0}$. What we are doing here is to assume that there is
some external source present such that we have the expectation value
specified. It follows that one of the expected results of our calculations
is the determination, at least implicitly, of the form of the external
source in terms of $v_{0}$.

Our first calculational task in preparation for the Gaussian-Perturbative
calculations is to rewrite the action in terms of a shifted field, which
has a null expectation value. We thus define a new field variable
$\vec{\varphi}'(n_{\mu})$ such that

\begin{displaymath}
  \vec{\varphi}(n_{\mu})
  =
  \vec{\varphi}'(n_{\mu})
  +
  (0,0,\ldots,v_{0}),
\end{displaymath}

\noindent
so that we have $\left\langle\vec{\varphi}'(n_{\mu})\right\rangle=0$ for
all $n_{\mu}$, with $\mu=1,\ldots,d$. We must now determine the form of
the action in terms of $\vec{\varphi}'(n_{\mu})$. If we write each term of
the action in terms of the shifted field we get

\noindent
\begin{eqnarray*}
  S_{N}[\vec{\varphi}']
  & = &
  \sum_{n_{\mu}}^{N^{d}}
  \left\{
    \rule{0em}{4ex}
    \frac{1}{2}
    \sum_{\nu}^{d}
    \left[
      \Delta_{\nu}\vec{\varphi}'(n_{\mu})
      \cdot
      \Delta_{\nu}\vec{\varphi}'(n_{\mu})
    \right]
    +
  \right.
  \\
  &   &
  \hspace{2.0em}
  \left.
    \rule{0em}{4ex}
    +
    \frac{\alpha}{2}
    \left[\vec{\varphi}'(n_{\mu})\cdot\vec{\varphi}'(n_{\mu})\right]
    +
    \alpha
    v_{0}\varphi_{\sN}'(n_{\mu})
    +
    \frac{\alpha}{2}\,
    v_{0}^{2}
    +
  \right.
  \\
  &   &
  \hspace{2.0em}
  \left.
    \rule{0em}{4ex}
    +
    \frac{\lambda}{4}
    \left[\vec{\varphi}'(n_{\mu})\cdot\vec{\varphi}'(n_{\mu})\right]^{2}
    +
    \lambda
    v_{0}
    \left[\vec{\varphi}'(n_{\mu})\cdot\vec{\varphi}'(n_{\mu})\right]
    \varphi_{\sN}'(n_{\mu})
    +
  \right.
  \\
  &   &
  \hspace{2.0em}
  \left.
    \rule{0em}{4ex}
    +
    \frac{\lambda}{2}\,
    v_{0}^{2}
    \left[\vec{\varphi}'(n_{\mu})\cdot\vec{\varphi}'(n_{\mu})\right]
    +
    \lambda
    v_{0}^{2}
    \varphi_{\sN}'^{2}(n_{\mu})
    +
  \right.
  \\
  &   &
  \hspace{2.0em}
  \left.
    \rule{0em}{4ex}
    +
    \lambda
    v_{0}^{3}\varphi_{\sN}'(n_{\mu})
    +
    \frac{\lambda}{4}\,
    v_{0}^{4}
    +
  \right.
  \\
  &   &
  \hspace{2.0em}
  \left.
    \rule{0em}{4ex}
    -
    j_{0}\varphi_{\sN}'(n_{\mu})
    -
    j_{0}v_{0}
  \right\}.
\end{eqnarray*}

\noindent
We will now eliminate all field-independent terms, since they correspond
to constant factors that cancel off in the ratios of functional integrals
which give the expectation values of the observables. Doing this we get
the equivalent action

\noindent
\begin{eqnarray*}
  S_{N}[\vec{\varphi}']
  & = &
  \sum_{n_{\mu}}^{N^{d}}
  \left\{
    \rule{0em}{4ex}
    \frac{1}{2}
    \sum_{\nu}^{d}
    \left[
      \Delta_{\nu}\vec{\varphi}'(n_{\mu})
      \cdot
      \Delta_{\nu}\vec{\varphi}'(n_{\mu})
    \right]
    +
  \right.
  \\
  &   &
  \hspace{2.0em}
  \left.
    \rule{0em}{4ex}
    +
    \alpha
    v_{0}\varphi_{\sN}'(n_{\mu})
    +
    \lambda
    v_{0}^{3}\varphi_{\sN}'(n_{\mu})
    -
    j_{0}\varphi_{\sN}'(n_{\mu})
    +
  \right.
  \\
  &   &
  \hspace{2.0em}
  \left.
    \rule{0em}{5.5ex}
    +
    \frac{\alpha+\lambda v_{0}^{2}}{2}
    \left[\vec{\varphi}'(n_{\mu})\cdot\vec{\varphi}'(n_{\mu})\right]
    +
    \lambda
    v_{0}^{2}
    \varphi_{\sN}'^{2}(n_{\mu})
    +
  \right.
  \\
  &   &
  \hspace{2.0em}
  \left.
    \rule{0em}{4ex}
    +
    \lambda
    v_{0}
    \left[\vec{\varphi}'(n_{\mu})\cdot\vec{\varphi}'(n_{\mu})\right]
    \varphi_{\sN}'(n_{\mu})
    +
    \frac{\lambda}{4}
    \left[\vec{\varphi}'(n_{\mu})\cdot\vec{\varphi}'(n_{\mu})\right]^{2}
  \right\},
\end{eqnarray*}

\noindent
where except for the kinetic part the terms have been ordered by
increasing powers of the field.

The last task we have to perform, in preparation for the
Gaussian-Perturbative calculations, is the separation of the action in two
parts. Since the symmetry is broken by the introduction of the external
sources, besides the fact that depending on the values of the parameters
$\alpha$ and $\lambda$ it might be spontaneously broken as well, this
separation involves two new mass parameters, $\alpha_{0}$ for
$\varphi_{1}'(n_{\mu}),\ldots,\varphi_{\sN-1}'(n_{\mu})$, and
$\alpha_{\sN}$ for $\varphi_{\sN}'(n_{\mu})$. Note that an $SO(\sN-1)$
symmetry subgroup is left over after the $SO(\sN)$ symmetry breakdown. We
therefore adopt for the Gaussian part of the action

\noindent
\begin{eqnarray}\label{ResultS0}
  S_{0}[\vec{\varphi}']
  & = &
  \sum_{n_{\mu}}^{N^{d}}
  \left\{
    \rule{0em}{4ex}
    \frac{1}{2}
    \sum_{\nu}^{d}
    \left[
      \Delta_{\nu}\vec{\varphi}'(n_{\mu})
      \cdot
      \Delta_{\nu}\vec{\varphi}'(n_{\mu})
    \right]
    +
  \right.
  \nonumber\\
  &   &
  \hspace{2.0em}
  \left.
    \rule{0em}{4ex}
    +
    \frac{\alpha_{0}}{2}
    \left[\vec{\varphi}'(n_{\mu})\cdot\vec{\varphi}'(n_{\mu})\right]
    +
    \frac{\alpha_{\sN}-\alpha_{0}}{2}\,
    \varphi_{\sN}'^{2}(n_{\mu})
  \right\},
\end{eqnarray}

\noindent
where there are no constraints on the parameters introduced other than
$\alpha_{0}\geq 0$ and $\alpha_{\sN}\geq 0$. Note that, despite the way in
which this is written, we do in fact have here just an $\alpha_{0}$ mass
term for each field component $\varphi_{i}'(n_{\mu})$, for
$i=1,\ldots,\sN-1$, and an $\alpha_{\sN}$ mass term for the field
component $\varphi_{\sN}'(n_{\mu})$. It follows that the non-Gaussian part
of the action is

\noindent
\begin{eqnarray}\label{ResultSV}
  S_{V}[\vec{\varphi}']
  & = &
  \sum_{n_{\mu}}^{N^{d}}
  \left\{
    \rule{0em}{3ex}
    v_{0}
    \left[\alpha+\lambda v_{0}^{2}\right]
    \varphi_{\sN}'(n_{\mu})
    -
    j_{0}\varphi_{\sN}'(n_{\mu})
    +
  \right.
  \nonumber\\
  &   &
  \hspace{2.23em}
  \left.
    +
    \frac{\alpha-\alpha_{0}+\lambda v_{0}^{2}}{2}
    \left[\vec{\varphi}'(n_{\mu})\cdot\vec{\varphi}'(n_{\mu})\right]
    +
    \frac{\alpha_{0}-\alpha_{\sN}+2\lambda v_{0}^{2}}{2}\,
    \varphi_{\sN}'^{2}(n_{\mu})
    +
  \right.
  \nonumber\\
  &   &
  \hspace{2.0em}
  \left.
    \rule{0em}{3ex}
    +
    \lambda
    v_{0}
    \left[\vec{\varphi}'(n_{\mu})\cdot\vec{\varphi}'(n_{\mu})\right]
    \varphi_{\sN}'(n_{\mu})
    +
    \frac{\lambda}{4}
    \left[\vec{\varphi}'(n_{\mu})\cdot\vec{\varphi}'(n_{\mu})\right]^{2}
  \right\},
\end{eqnarray}

\noindent
which has its terms now written strictly in the order of increasing powers
of the field.

Let us end this section by recalling the calculational techniques that
will be involved. Given an arbitrary observable $\OO[\vec{\varphi}']$ its
expectation value is defined by

\begin{displaymath}
  \left\langle\OO[\vec{\varphi}']\right\rangle
  =
  \frac
  {
    \displaystyle
    \int[{\rm d}\varphi]\,
    \OO[\vec{\varphi}']\,
    e^{-S_{0}[\vec{\varphi}']-\xi S_{V}[\vec{\varphi}']}
  }
  {
    \displaystyle
    \int[{\rm d}\varphi]\,
    e^{-S_{0}[\vec{\varphi}']-\xi S_{V}[\vec{\varphi}']}
  },
\end{displaymath}

\noindent
which is a function of $\xi$, where $[{\rm d}\varphi]$ denotes the flat
measure and hence integrals from $-\infty$ to $+\infty$ over all the field
components at all sites. The expectation values of the model are obtained
for $\xi=1$, and the corresponding expectation values in the Gaussian
measure of $S_{0}[\vec{\varphi}']$ are those obtained for $\xi=0$. The
Gaussian-Perturbative approximation consists of the expansion of the
right-hand side in powers of $\xi$ to some finite order, around the point
$\xi=0$, and the application of the resulting expression at $\xi=1$. The
first-order Gaussian-Perturbative approximation of the expectation value
of the observable $\OO[\vec{\varphi}']$ is given by

\noindent
\begin{eqnarray*}
  \left\langle\OO[\vec{\varphi}']\right\rangle
  & = &
  \left\langle\OO[\vec{\varphi}']\right\rangle_{0}
  -
  \left\{
    \left\langle\OO[\vec{\varphi}']S_{V}[\vec{\varphi}']\right\rangle_{0}
    -
    \left\langle\OO[\vec{\varphi}']\right\rangle_{0}
    \left\langle S_{V}[\vec{\varphi}']\right\rangle_{0}
  \right\},
\end{eqnarray*}

\noindent
where the subscript $0$ indicates the expectation values in the measure of
$S_{0}[\vec{\varphi}']$. These expectation values are most easily
calculated in momentum space, where they involve only uncoupled Gaussian
integrals. Therefore, let us also recall here the transformations to and
from the momentum space representation of the model. We have for the field
and its Fourier transform $\tvphi_{i}'(k_{\mu})$

\noindent
\begin{eqnarray*}
  \tvphi_{i}'(k_{\mu})
  & = &
  \frac{1}{N^{d}}\sum_{n_{\mu}}^{N^{d}}
  e^{\ii(2\pi/N)\sum_{\mu}^{d}k_{\mu}n_{\mu}}\,\varphi_{i}'(n_{\mu}),
  \\
  \varphi_{i}'(n_{\mu})
  & = &
  \sum_{k_{\mu}}^{N^{d}}
  e^{-\ii(2\pi/N)\sum_{\mu}^{d}k_{\mu}n_{\mu}}\,\tvphi_{i}'(k_{\mu}),
\end{eqnarray*}

\noindent
where the sums over $k_{\mu}$ are taken in as symmetric a way as possible
around $k_{\mu}=0_{\mu}$, just as we did for $n_{\mu}$. In other words, we
have $k_{\mu}=k_{\rm min},\ldots,0,\ldots,k_{\rm max}$ with the same
values of $k_{\rm min}$ and $k_{\rm max}$, depending on the parity of $N$,
that were used for $n_{\rm min}$ and $n_{\rm max}$,

\begin{displaymath}
  k_{\mu}
  =
  -\,\frac{N-1}{2},\ldots,0,\ldots,\frac{N-1}{2},
\end{displaymath}

\noindent
for odd $N$, and

\begin{displaymath}
  k_{\mu}
  =
  -\,\frac{N}{2}+1,\ldots,0,\ldots,\frac{N}{2},
\end{displaymath}

\noindent
for even $N$, in either case for all values of $\mu=1,\ldots,d$. The
orthogonality and completeness relations of the Fourier base are given by

\noindent
\begin{eqnarray*}
  \sum_{n_{\mu}}^{N^{d}}
  e^{\pm\ii(2\pi/N)\sum_{\mu}^{d}n_{\mu}(k_{\mu}-k_{\mu}')}
  & = &
  N^{d}\delta^{d}(k_{\mu},k_{\mu}'),
  \\
  \sum_{k_{\mu}}^{N^{d}}
  e^{\pm\ii(2\pi/N)\sum_{\mu}^{d}k_{\mu}(n_{\mu}-n_{\mu}')}
  & = &
  N^{d}\delta^{d}(n_{\mu},n_{\mu}').
\end{eqnarray*}

\noindent
A typical Gaussian expectation value in momentum space, and possibly the
most fundamental one, is given for a generic field component by

\begin{displaymath}
  \left\langle
    \tvphi_{i}'(k_{\mu})
    \tvphi_{i}'^{*}(k_{\mu})
  \right\rangle_{0}
  =
  \frac{1}{N^{d}}\,
  \frac{1}{\rho^{2}(k_{\mu})+\alpha_{i}},
\end{displaymath}

\noindent
where $\alpha_{i}$ is either $\alpha_{0}$ or $\alpha_{\sN}$, depending on
the field component involved, and where $\rho^{2}(k_{\mu})$ are the
eigenvalues of the discrete Laplacian on the lattice, which are given by

\begin{displaymath}
  \rho^{2}(k_{\mu})
  =
  4
  \left[
    \sin^{2}\!\left(\frac{\pi k_{1}}{N}\right)
    +
    \ldots
    +
    \sin^{2}\!\left(\frac{\pi k_{d}}{N}\right)
  \right].
\end{displaymath}

\noindent
This and several other expectation values, Gaussian integration formulas
and lattice sums can be found in Appendix~\ref{TabofInts}.

\section{Calculations}

We are now ready for the Gaussian-Perturbative calculations. We start with
the calculations involving the critical behavior of the model, so that we
may determine and characterize its two phases. It is important to observe
that the phase structure of the model must be established right at the
beginning, because everything else has to be discussed in terms of it.

\subsection{The Critical Line}

We will now calculate the Gaussian-Perturbative approximation for the
particular observable $\OO[\vec{\varphi}']=\varphi_{\sN}(n_{\mu}')$, at
some arbitrary point $n_{\mu}'$. If we write the observable in terms of
the shifted field we get

\noindent
\begin{eqnarray*}
  \OO[\vec{\varphi}']
  & = &
  \varphi_{\sN}(n_{\mu}')
  \\
  & = &
  \varphi_{\sN}'(n_{\mu}')+v_{0}.
\end{eqnarray*}

\noindent
In order to get the equation of the critical line we impose, in a
self-consistent way, that we in fact have

\begin{displaymath}
  \left\langle\varphi_{\sN}(n_{\mu}')\right\rangle
  =
  v_{0},
\end{displaymath}

\noindent
which is the same as stating that

\begin{displaymath}
  \left\langle\varphi_{\sN}'(n_{\mu}')\right\rangle
  =
  0.
\end{displaymath}

\noindent
In the first-order Gaussian-Perturbative approximation this becomes

\noindent
\begin{eqnarray*}
  \left\langle\varphi_{\sN}'(n_{\mu}')\right\rangle
  & = &
  \left\langle\varphi_{\sN}'(n_{\mu}')\right\rangle_{0}
  -
  \left\{
    \left\langle
      \varphi_{\sN}'(n_{\mu}')S_{V}[\vec{\varphi}']
    \right\rangle_{0}
    -
    \left\langle\varphi_{\sN}'(n_{\mu}')\right\rangle_{0}
    \left\langle
      S_{V}[\vec{\varphi}']
    \right\rangle_{0}
  \right\}
  \\
  & = &
  0.
\end{eqnarray*}

\noindent
Since we have $\left\langle\varphi_{\sN}'(n_{\mu}')\right\rangle_{0}=0$,
because this observable is field-odd and the Gaussian action
$S_{0}[\vec{\varphi}']$ is field-even, we get for the critical line the
simple equation, known as the tadpole equation,

\begin{displaymath}
  \left\langle
    \varphi_{\sN}'(n_{\mu}')S_{V}[\vec{\varphi}']
  \right\rangle_{0}
  =
  0.
\end{displaymath}

\noindent
The expectation value shown here is calculated in
Appendix~\ref{ExpecVals}, given in Equation~(\ref{ExpecValFNSV}), and the
result is

\begin{displaymath}
  \left\langle
    \varphi_{\sN}'(n_{\mu}')S_{V}[\vec{\varphi}']
  \right\rangle_{0}
  =
  \frac
  {
    v_{0}
    \left[
      \alpha
      +
      v_{0}^{2}
      \lambda
      +
      (\sN-1)
      \lambda
      \sigma_{0}^{2}
      +
      3\lambda
      \sigma_{\sN}^{2}
    \right]
    -
    j_{0}
  }
  {\alpha_{\sN}}.
\end{displaymath}

\noindent
The parameter $\alpha_{\sN}$ cancels off from our equation, and thus we
are left with the result

\begin{equation}\label{ResultCL}
  j_{0}
  =
  v_{0}
  \left\{
    \lambda
    v_{0}^{2}
    +
    \alpha
    +
    \lambda
    \left[
      (\sN-1)
      \sigma_{0}^{2}
      +
      3
      \sigma_{\sN}^{2}
    \right]
  \right\},
\end{equation}

\noindent
in which we now isolated on the left-hand side the term with the external
source. This gives the general relation between $j_{0}$ and $v_{0}$ at
each point $(\alpha,\lambda)$ of the parameter space of the model. As we
shall see later, from this result we can determine the critical behavior
of the model and derive the equation of the critical line.

The quantity $\sigma_{0}$ is the width or variance of the local
distribution of values of the field components $\varphi_{i}'(n_{\mu})$,
with $i=1,\ldots,\sN-1$, in the measure of $S_{0}[\vec{\varphi}']$,

\noindent
\begin{eqnarray*}
  \sigma_{0}^{2}
  & = &
  \left\langle
    \varphi_{i}'^{2}(n_{\mu})
  \right\rangle_{0},
  \\
  & = &
  \frac{1}{N^{d}}
  \sum_{k_{\mu}}^{N^{d}}
  \frac{1}{\rho^{2}(k_{\mu})+\alpha_{0}},
\end{eqnarray*}

\noindent
as one can see in Appendix~\ref{TabofInts}, Equation~(\ref{Sigma0Form}),
and has the following interesting properties, so long as $d\geq 3$. First,
it is independent of the position $n_{\mu}$, as translation invariance
would require. Second, for $d\geq 3$ it has a finite and non-zero
$N\to\infty$ limit, so long as $\alpha_{0}=a^{2}m_{0}^{2}$ with a finite
value of $m_{0}$ in the limit. Finally, the value of $\sigma_{0}$ in the
limit does not depend on the value of $m_{0}$ in that same limit.
Analogously, the quantity $\sigma_{\sN}$ is associated to the remaining
field component $\varphi_{\sN}'(n_{\mu})$ and to the mass parameter
$\alpha_{\sN}$, and has these same properties. In fact, $\sigma_{0}$ and
$\sigma_{\sN}$ have exactly the same value in the $N\to\infty$ limit.

\subsection{The Transversal Propagator}

We will now calculate the expectation value of the observable

\begin{displaymath}
  \OO[\vec{\varphi}']
  =
  \varphi_{i}'(n_{\mu}')
  \varphi_{i}'(n_{\mu}''),
\end{displaymath}

\noindent
which has the same form for all components of the field except $i=\sN$.
We call this the transversal propagator because it belongs to the field
components which are orthogonal to the direction of the external source in
the internal $SO(\sN)$ space. In this section we will assume that
$i\neq\sN$, in fact we will make $i=1$. The observable will be taken at
two arbitrary points $n_{\mu}'$ and $n_{\mu}''$. The first-order
Gaussian-Perturbative approximation for this observable gives

\noindent
\begin{eqnarray*}
  \left\langle
    \varphi_{1}'(n_{\mu}')
    \varphi_{1}'(n_{\mu}'')
  \right\rangle
  & = &
  \left\langle
    \varphi_{1}'(n_{\mu}')
    \varphi_{1}'(n_{\mu}'')
  \right\rangle_{0}
  +
  \\
  &   &
  -
  \left\{
    \left\langle
      \varphi_{1}'(n_{\mu}')
      \varphi_{1}'(n_{\mu}'')
      S_{V}[\vec{\varphi}']
    \right\rangle_{0}
    -
    \left\langle
      \varphi_{1}'(n_{\mu}')
      \varphi_{1}'(n_{\mu}'')
    \right\rangle_{0}
    \left\langle
      S_{V}[\vec{\varphi}']
    \right\rangle_{0}
  \right\}
  \\
  & = &
  g_{0}(n_{\mu}'-n_{\mu}'')
  +
  \\
  &   &
  -
  \left\{
    \left\langle
      \varphi_{1}'(n_{\mu}')
      \varphi_{1}'(n_{\mu}'')
      S_{V}[\vec{\varphi}']
    \right\rangle_{0}
    -
    g_{0}(n_{\mu}'-n_{\mu}'')
    \left\langle
      S_{V}[\vec{\varphi}']
    \right\rangle_{0}
  \right\},
\end{eqnarray*}

\noindent
where $g_{0}(n_{\mu}'-n_{\mu}'')$ is the two-point function with mass
parameter $\alpha_{0}$. We must calculate the two expectation values which
appear in this formula. The calculation of the first one is done in
Appendix~\ref{ExpecVals}, given in Equation~(\ref{ExpecValSV}), and
results in

\noindent
\begin{eqnarray*}
  \left\langle
    S_{V}[\vec{\varphi}']
  \right\rangle_{0}
  & = &
  N^{d}
  \left[
    \frac{\alpha-\alpha_{0}+\lambda v_{0}^{2}}{2}\,
    (\sN-1)
    \sigma_{0}^{2}
    +
    \frac{\alpha-\alpha_{\sN}+3\lambda v_{0}^{2}}{2}\,
    \sigma_{\sN}^{2}
    +
  \right.
  \\
  &   &
  \hspace{2.2em}
  \left.
    +
    \frac{\lambda}{4}\,
    (\sN^{2}-1)
    \sigma_{0}^{4}
    +
    \frac{\lambda}{2}\,
    (\sN-1)
    \sigma_{0}^{2}
    \sigma_{\sN}^{2}
    +
    \frac{3\lambda}{4}\,
    \sigma_{\sN}^{4}
  \right].
\end{eqnarray*}

\noindent
The second expectation value is also calculated in
Appendix~\ref{ExpecVals}, given in Equation~(\ref{ExpecValF1F1SV}), and
the result is

\noindent
\begin{eqnarray*}
  \lefteqn
  {
    \left\langle
      \varphi_{1}'(n_{\mu}')
      \varphi_{1}'(n_{\mu}'')
      S_{V}[\vec{\varphi}']
    \right\rangle_{0}
  }
  &   &
  \\
  & = &
  N^{d}
  \left[
    \frac{\alpha-\alpha_{0}+\lambda v_{0}^{2}}{2}\,
    (\sN-1)
    \sigma_{0}^{2}
    +
    \frac{\alpha-\alpha_{\sN}+3\lambda v_{0}^{2}}{2}\,
    \sigma_{\sN}^{2}
    +
  \right.
  \\
  &   &
  \hspace{2.2em}
  \left.
    +
    \frac{\lambda}{4}\,
    (\sN^{2}-1)
    \sigma_{0}^{4}
    +
    \frac{\lambda}{2}\,
    (\sN-1)
    \sigma_{0}^{2}
    \sigma_{\sN}^{2}
    +
    \frac{3\lambda}{4}\,
    \sigma_{\sN}^{4}
  \right]
  g_{0}(n_{\mu}'-n_{\mu}'')
  \\
  &   &
  +
  \left[
    \alpha-\alpha_{0}
    +
    \lambda
    v_{0}^{2}
    +
    \lambda
    (\sN+1)
    \sigma_{0}^{2}
    +
    \lambda
    \sigma_{\sN}^{2}
  \right]
  \frac{1}{N^{d}}
  \sum_{k_{\mu}}^{N^{d}}
  \frac
  {
    e^{-\ii(2\pi/N)\sum_{\mu}^{d}k_{\mu}(n_{\mu}'-n_{\mu}'')}
  }
  {
    [\rho^{2}(k_{\mu})+\alpha_{0}]^{2}
  }.
\end{eqnarray*}

\noindent
The factor in front of $g_{0}(n_{\mu}'-n_{\mu}'')$ can now be verified to
be exactly equal to $\left\langle S_{V}[\vec{\varphi}']\right\rangle_{0}$,
and therefore this whole part cancels off from our observable. We may now
write for the difference of expectation values that appears in it,

\noindent
\begin{eqnarray*}
  \lefteqn
  {
    \left\langle
      \varphi_{1}'(n_{\mu}')
      \varphi_{1}'(n_{\mu}'')
      S_{V}[\vec{\varphi}']
    \right\rangle_{0}
    -
    \left\langle
      S_{V}[\vec{\varphi}']
    \right\rangle_{0}
    g_{0}(n_{\mu}'-n_{\mu}'')
  }
  &   &
  \\
  & = &
  \left[
    \lambda
    v_{0}^{2}
    +
    \alpha-\alpha_{0}
    +
    \lambda
    (\sN+1)
    \sigma_{0}^{2}
    +
    \lambda
    \sigma_{\sN}^{2}
  \right]
  \frac{1}{N^{d}}
  \sum_{k_{\mu}}^{N^{d}}
  \frac
  {
    e^{-\ii(2\pi/N)\sum_{\mu}^{d}k_{\mu}(n_{\mu}'-n_{\mu}'')}
  }
  {
    [\rho^{2}(k_{\mu})+\alpha_{0}]^{2}
  }.
\end{eqnarray*}

\noindent
Finally, we can write the complete result,

\noindent
\begin{eqnarray*}
  \lefteqn
  {
    \left\langle
      \varphi_{1}'(n_{\mu}')
      \varphi_{1}'(n_{\mu}'')
    \right\rangle
  }
  &   &
  \\
  & = &
  \frac{1}{N^{d}}
  \sum_{k_{\mu}}^{N^{d}}
  \frac
  {
    e^{-\ii(2\pi/N)\sum_{\mu}^{d}k_{\mu}(n_{\mu}'-n_{\mu}'')}
  }
  {
    \rho^{2}(k_{\mu})+\alpha_{0}
  }
  +
  \\
  &   &
  -
  \left[
    \lambda
    v_{0}^{2}
    +
    \alpha-\alpha_{0}
    +
    \lambda
    (\sN+1)
    \sigma_{0}^{2}
    +
    \lambda
    \sigma_{\sN}^{2}
  \right]
  \frac{1}{N^{d}}
  \sum_{k_{\mu}}^{N^{d}}
  \frac
  {
    e^{-\ii(2\pi/N)\sum_{\mu}^{d}k_{\mu}(n_{\mu}'-n_{\mu}'')}
  }
  {
    [\rho^{2}(k_{\mu})+\alpha_{0}]^{2}
  }
  \\
  & = &
  \frac{1}{N^{d}}
  \sum_{k_{\mu}}^{N^{d}}
  \frac
  {
    e^{-\ii(2\pi/N)\sum_{\mu}^{d}k_{\mu}(n_{\mu}'-n_{\mu}'')}
  }
  {
    \rho^{2}(k_{\mu})+\alpha_{0}
  }
  \left[
    1
    -
    \frac
    {
      \lambda
      v_{0}^{2}
      +
      \alpha-\alpha_{0}
      +
      \lambda
      (\sN+1)
      \sigma_{0}^{2}
      +
      \lambda
      \sigma_{\sN}^{2}
    }
    {
      \rho^{2}(k_{\mu})+\alpha_{0}
    }
  \right],
\end{eqnarray*}

\noindent
where we wrote $g_{0}(n_{\mu}'-n_{\mu}'')$ in terms of its Fourier
transform.

In principle we could have used any positive value of $\alpha_{0}$ for
this calculation, but now a particular choice comes to our attention. We
see from the structure of this propagator that we can make $\alpha_{0}$
equal to the transversal renormalized mass parameter by choosing it so
that the numerator of the second fraction vanishes. In this way we get a
very simple propagator, with a simple pole in the complex $\rho^{2}$
plane, in which the parameter $\alpha_{0}$ appears now in the role of the
renormalized mass parameter,

\begin{displaymath}
  \left\langle
    \varphi_{1}'(n_{\mu}')
    \varphi_{1}'(n_{\mu}'')
  \right\rangle
  =
  \frac{1}{N^{d}}
  \sum_{k_{\mu}}^{N^{d}}
  \frac
  {
    e^{-\ii(2\pi/N)\sum_{\mu}^{d}k_{\mu}(n_{\mu}'-n_{\mu}'')}
  }
  {
    \rho^{2}(k_{\mu})+\alpha_{0}
  }.
\end{displaymath}

\noindent
Observe that to this order the propagator is, in fact, the propagator of
the free theory. This is a self-consistent way to choose the parameter
$\alpha_{0}$, and is equivalent to the determination of the transversal
renormalized mass. This choice is equivalent to requiring that the mass
parameter of the Gaussian measure being used for the approximation of the
expectation values be the same as the renormalized mass parameter of the
original quantum model. It gives the result

\begin{equation}\label{ResultAlpha0}
  \alpha_{0}
  =
  \lambda
  v_{0}^{2}
  +
  \alpha
  +
  \lambda
  \left[
    (\sN+1)
    \sigma_{0}^{2}
    +
    \sigma_{\sN}^{2}
  \right].
\end{equation}

\noindent
This result for $\alpha_{0}=a^{2}m_{0}^{2}$ is valid for a constant but
possibly non-zero external source, in both phases of the model, where
$m_{0}$ is the mass associated to the $\sN-1$ field components
$\varphi_{i}'(n_{\mu})$, for $i\neq\sN$.

\subsection{The Longitudinal Propagator}

We will now complete our calculations with the expectation value of the
observable

\begin{displaymath}
  \OO[\vec{\varphi}']
  =
  \varphi_{\sN}'(n_{\mu}')
  \varphi_{\sN}'(n_{\mu}'').
\end{displaymath}

\noindent
We call this the longitudinal propagator because it belongs to the field
component which is in the direction of the external source in the internal
$SO(\sN)$ space. Once more the observable will be taken at two arbitrary
points $n_{\mu}'$ and $n_{\mu}''$. The first-order Gaussian-Perturbative
approximation for this observable gives

\noindent
\begin{eqnarray*}
  \left\langle
    \varphi_{\sN}'(n_{\mu}')
    \varphi_{\sN}'(n_{\mu}'')
  \right\rangle
  & = &
  \left\langle
    \varphi_{\sN}'(n_{\mu}')
    \varphi_{\sN}'(n_{\mu}'')
  \right\rangle_{0}
  +
  \\
  &   &
  -
  \left\{
    \left\langle
      \varphi_{\sN}'(n_{\mu}')
      \varphi_{\sN}'(n_{\mu}'')
      S_{V}[\vec{\varphi}']
    \right\rangle_{0}
    -
    \left\langle
      \varphi_{\sN}'(n_{\mu}')
      \varphi_{\sN}'(n_{\mu}'')
    \right\rangle_{0}
    \left\langle
      S_{V}[\vec{\varphi}']
    \right\rangle_{0}
  \right\}
  \\
  & = &
  g_{\sN}(n_{\mu}'-n_{\mu}'')
  +
  \\
  &   &
  -
  \left\{
    \left\langle
      \varphi_{\sN}'(n_{\mu}')
      \varphi_{\sN}'(n_{\mu}'')
      S_{V}[\vec{\varphi}']
    \right\rangle_{0}
    -
    g_{\sN}(n_{\mu}'-n_{\mu}'')
    \left\langle
      S_{V}[\vec{\varphi}']
    \right\rangle_{0}
  \right\},
\end{eqnarray*}

\noindent
where $g_{\sN}(n_{\mu}'-n_{\mu}'')$ is the two-point function with mass
parameter $\alpha_{\sN}$. We must now calculate the two expectation values
which appear in this formula. The first expectation value is the same we
had before for the transversal propagator, and from
Equation~(\ref{ExpecValSV}) we have,

\noindent
\begin{eqnarray*}
  \left\langle
    S_{V}[\vec{\varphi}']
  \right\rangle_{0}
  & = &
  N^{d}
  \left[
    \frac{\alpha-\alpha_{0}+\lambda v_{0}^{2}}{2}\,
    (\sN-1)
    \sigma_{0}^{2}
    +
    \frac{\alpha-\alpha_{\sN}+3\lambda v_{0}^{2}}{2}\,
    \sigma_{\sN}^{2}
    +
  \right.
  \\
  &   &
  \hspace{2.2em}
  \left.
    +
    \frac{\lambda}{4}\,
    (\sN^{2}-1)
    \sigma_{0}^{4}
    +
    \frac{\lambda}{2}\,
    (\sN-1)
    \sigma_{0}^{2}
    \sigma_{\sN}^{2}
    +
    \frac{3\lambda}{4}\,
    \sigma_{\sN}^{4}
  \right].
\end{eqnarray*}

\noindent
The second expectation value is calculated in Appendix~\ref{ExpecVals},
given in Equation~(\ref{ExpecValFNFNSV}),and the result is

\noindent
\begin{eqnarray*}
  \lefteqn
  {
    \left\langle
      \varphi_{\sN}'(n_{\mu}')
      \varphi_{\sN}'(n_{\mu}'')
      S_{V}[\vec{\varphi}']
    \right\rangle_{0}
  }
  &   &
  \\
  & = &
  N^{d}
  \left[
    \frac{\alpha-\alpha_{0}+\lambda v_{0}^{2}}{2}\,
    (\sN-1)
    \sigma_{0}^{2}
    +
    \frac{\alpha-\alpha_{\sN}+3\lambda v_{0}^{2}}{2}\,
    \sigma_{\sN}^{2}
    +
  \right.
  \\
  &   &
  \hspace{2.2em}
  \left.
    +
    \frac{\lambda}{4}\,
    (\sN^{2}-1)
    \sigma_{0}^{4}
    +
    \frac{\lambda}{2}\,
    (\sN-1)
    \sigma_{0}^{2}
    \sigma_{\sN}^{2}
    +
    \frac{3\lambda}{4}\,
    \sigma_{\sN}^{4}
  \right]
  g_{\sN}(n_{\mu}'-n_{\mu}'')
  \\
  &   &
  +
  \left[
    \alpha-\alpha_{\sN}
    +
    3\lambda
    v_{0}^{2}
    +
    \lambda
    (\sN-1)
    \sigma_{0}^{2}
    +
    3\lambda
    \sigma_{\sN}^{2}
  \right]
  \frac{1}{N^{d}}
  \sum_{k_{\mu}}^{N^{d}}
  \frac
  {
    e^{-\ii(2\pi/N)\sum_{\mu}^{d}k_{\mu}(n_{\mu}'-n_{\mu}'')}
  }
  {
    [\rho^{2}(k_{\mu})+\alpha_{\sN}]^{2}
  }.
\end{eqnarray*}

\noindent
The factor in front of $g_{\sN}(n_{\mu}'-n_{\mu}'')$ can now be verified
to be exactly equal to $\left\langle
  S_{V}[\vec{\varphi}']\right\rangle_{0}$, and therefore once again this
whole part cancels off from our observable. We may now write for the
difference of expectation values that appears in it,

\noindent
\begin{eqnarray*}
  \lefteqn
  {
    \left\langle
      \varphi_{\sN}'(n_{\mu}')
      \varphi_{\sN}'(n_{\mu}'')
      S_{V}[\vec{\varphi}']
    \right\rangle_{0}
    -
    \left\langle
      S_{V}[\vec{\varphi}']
    \right\rangle_{0}
    g_{\sN}(n_{\mu}'-n_{\mu}'')
  }
  &   &
  \\
  & = &
  \left[
    3\lambda
    v_{0}^{2}
    +
    \alpha-\alpha_{\sN}
    +
    \lambda
    (\sN-1)
    \sigma_{0}^{2}
    +
    3\lambda
    \sigma_{\sN}^{2}
  \right]
  \frac{1}{N^{d}}
  \sum_{k_{\mu}}^{N^{d}}
  \frac
  {
    e^{-\ii(2\pi/N)\sum_{\mu}^{d}k_{\mu}(n_{\mu}'-n_{\mu}'')}
  }
  {
    [\rho^{2}(k_{\mu})+\alpha_{\sN}]^{2}
  }.
\end{eqnarray*}

\noindent
Finally, we can write the complete result,

\noindent
\begin{eqnarray*}
  \lefteqn
  {
    \left\langle
      \varphi_{\sN}'(n_{\mu}')
      \varphi_{\sN}'(n_{\mu}'')
    \right\rangle
  }
  &   &
  \\
  & = &
  \frac{1}{N^{d}}
  \sum_{k_{\mu}}^{N^{d}}
  \frac
  {
    e^{-\ii(2\pi/N)\sum_{\mu}^{d}k_{\mu}(n_{\mu}'-n_{\mu}'')}
  }
  {
    \rho^{2}(k_{\mu})+\alpha_{\sN}
  }
  +
  \\
  &   &
  -
  \left[
    3\lambda
    v_{0}^{2}
    +
    \alpha-\alpha_{\sN}
    +
    \lambda
    (\sN-1)
    \sigma_{0}^{2}
    +
    3\lambda
    \sigma_{\sN}^{2}
  \right]
  \frac{1}{N^{d}}
  \sum_{k_{\mu}}^{N^{d}}
  \frac
  {
    e^{-\ii(2\pi/N)\sum_{\mu}^{d}k_{\mu}(n_{\mu}'-n_{\mu}'')}
  }
  {
    [\rho^{2}(k_{\mu})+\alpha_{\sN}]^{2}
  }
  \\
  & = &
  \frac{1}{N^{d}}
  \sum_{k_{\mu}}^{N^{d}}
  \frac
  {
    e^{-\ii(2\pi/N)\sum_{\mu}^{d}k_{\mu}(n_{\mu}'-n_{\mu}'')}
  }
  {
    \rho^{2}(k_{\mu})+\alpha_{\sN}
  }
  \left[
    1
    -
    \frac
    {
      3\lambda
      v_{0}^{2}
      +
      \alpha-\alpha_{\sN}
      +
      \lambda
      (\sN-1)
      \sigma_{0}^{2}
      +
      3\lambda
      \sigma_{\sN}^{2}
    }
    {
      \rho^{2}(k_{\mu})+\alpha_{\sN}
    }
  \right],
\end{eqnarray*}

\noindent
where we once more wrote $g_{\sN}(n_{\mu}'-n_{\mu}'')$ in terms of its
Fourier transform. Exactly as in the previous case, we see from the
structure of this propagator that we can make $\alpha_{\sN}$ equal to the
longitudinal renormalized mass parameter by choosing it so that the
numerator of the second fraction vanishes. This gives the result

\begin{equation}\label{ResultAlphaN}
  \alpha_{\sN}
  =
  3\lambda
  v_{0}^{2}
  +
  \alpha
  +
  \lambda
  \left[
    (\sN-1)
    \sigma_{0}^{2}
    +
    3
    \sigma_{\sN}^{2}
  \right].
\end{equation}

\noindent
This result for $\alpha_{\sN}=a^{2}m_{\sN}^{2}$ is valid for a constant
but possibly non-zero external source, in both phases of the model, where
$m_{\sN}$ is the mass associated to the field component
$\varphi_{\sN}'(n_{\mu})$.

\section{Discussion}

In this section we analyze and discuss the physical significance of the
results obtained with the Gaussian-Perturbative approximation, starting
with the determination of the critical behavior of the model. As was
pointed out before, it is important that the phase structure of the model
be established right at the beginning, because everything else has to be
discussed in terms of it.

\subsection{Critical Behavior}

\paragraph{Critical Line:} Here we discuss the physical significance of
our result for $j_{0}$ as a function of $v_{0}$. As we shall see, this
bears on the critical behavior of the model. First of all, let us discuss
the case $j_{0}=0$, that is without external sources at all, which from
Equation~(\ref{ResultCL}) results in the equation

\begin{displaymath}
  v_{0}
  \left\{
    \lambda
    v_{0}^{2}
    +
    \alpha
    +
    \lambda
    \left[
      (\sN-1)
      \sigma_{0}^{2}
      +
      3
      \sigma_{\sN}^{2}
    \right]
  \right\}
  =
  0.
\end{displaymath}

\noindent
Observe that we do {\em not} assume that $v_{0}$ is automatically zero.
Since we must have $\lambda\geq 0$, in the part of the parameter space of
the model in which the quantity shown below is positive,

\begin{displaymath}
  \alpha
  +
  \lambda
  \left[
    (\sN-1)
    \sigma_{0}^{2}
    +
    3
    \sigma_{\sN}^{2}
  \right]
  >
  0,
\end{displaymath}

\noindent
the only possible solution to the equation is in fact $v_{0}=0$. This is
the {\em symmetric phase} of the model. On the other hand, in the
complementary region of the parameter space of the model, in which that
same quantity is negative,

\begin{displaymath}
  \alpha
  +
  \lambda
  \left[
    (\sN-1)
    \sigma_{0}^{2}
    +
    3
    \sigma_{\sN}^{2}
  \right]
  <
  0,
\end{displaymath}

\noindent
and once again because we must have $\lambda\geq 0$, there are two other
solutions besides the $v_{0}=0$ solution, given by

\begin{equation}\label{ResultvSSB}
  \lambda
  v_{0}^{2}
  =
  -
  \left\{
    \alpha
    +
    \lambda
    \left[
      (\sN-1)
      \sigma_{0}^{2}
      +
      3
      \sigma_{\sN}^{2}
    \right]
  \right\}.
\end{equation}

\noindent
Let us observe that since $\lambda>0$ we must have $\alpha<0$ here. This
is the {\em broken-symmetric phase} of the model, where these solutions
corresponds to the local minima of the potential, while $v_{0}=0$
corresponds to the local maximum. If we look for the locus in the
$(\alpha,\lambda)$ parameter plane of the model in which the $v_{0}=0$
solution becomes the {\em only} possible solution, we arrive at the
equation

\begin{displaymath}
  \alpha
  +
  \lambda
  \left[
    (\sN-1)
    \sigma_{0}^{2}
    +
    3
    \sigma_{\sN}^{2}
  \right]
  =
  0.
\end{displaymath}

\noindent
This is an equation giving $\lambda$ in terms of $\alpha$, which thus
determines a certain curve in the parameter plane of the model, in this
case a straight line. This is the {\em critical line}, which separates the
two phases of the model. An example of the parameter space of the model,
showing the critical line, can be seen in Figure~\ref{CritFig}, on
page~\pageref{CritFig}. To the right of this line the model is symmetric
and we have $\left\langle\vec{\varphi}(n_{\mu})\right\rangle=0$. To the
left, the symmetry is broken and we have
$\left\langle\varphi_{\sN}(n_{\mu})\right\rangle\neq 0$.

It might seem odd that we find here what looks like a complete phase
transition even on finite lattices. In fact, it is a known fact that there
are no real phase transitions on finite lattices with periodical boundary
conditions, a situation in which all one can hope to get are
approximations of this behavior. However, it is in fact possible to get
complete phase transitions on finite lattices if one uses other boundary
conditions or changes other aspects of the system, such as the imposition
of self-consistency conditions~\cite{fixfloat}, just as we do in the
Gaussian-Perturbative technique. Strictly speaking, however, the position
of the critical line that we find here is not completely well-defined on
finite lattices, because there is a slight dependence on $\alpha_{0}$ and
$\alpha_{\sN}$ through $\sigma_{0}$ and $\sigma_{\sN}$. This small
dependence vanishes in the continuum limit, of course.

Since $\sigma_{0}$ and $\sigma_{\sN}$ are strictly positive, and
$(\sN-1)\geq 0$, we can see that this critical line starts at the Gaussian
point $(\alpha,\lambda)=(0,0)$, and extends to the quadrant where
$\alpha<0$ and $\lambda>0$. Besides, since in the continuum limit
$\sigma_{0}$ and $\sigma_{\sN}$ become identical, we may write the
following equivalent equation for purposes of that limit,

\begin{equation}\label{CriticalLine}
  \alpha
  +
  (\sN+2)
  \lambda
  \sigma_{0}^{2}
  =
  0.
\end{equation}

\noindent
This is the known result for the critical line, obtained previously
without the introduction of any external sources at all~\cite{pertheory}.

Going back to the general case, when the external source $j_{0}$ is not
zero, then the equation of the critical line determines it in terms of
$v_{0}$, in either phase of the model, for as we saw before in
Equation~(\ref{ResultCL}) we have

\begin{displaymath}
  j_{0}
  =
  v_{0}
  \left\{
    \lambda
    v_{0}^{2}
    +
    \alpha
    +
    \lambda
    \left[
      (\sN-1)
      \sigma_{0}^{2}
      +
      3
      \sigma_{\sN}^{2}
    \right]
  \right\}.
\end{displaymath}

\noindent
Given a point $(\alpha,\lambda)$ in the parameter space of the model, this
clearly and directly determines $j_{0}$ in terms of $v_{0}$. Conversely,
given $j_{0}$ one may determine the corresponding $v_{0}$ by solving this
simple algebraic cubic equation. Using the result for $\alpha_{\sN}$ in
Equation~(\ref{ResultAlphaN}), we may write this cubic equation in a
simpler and more explicit form, in terms of the renormalized mass,

\begin{equation}\label{ResultCubicEq}
  v_{0}^{3}
  -
  \left(
    \frac{\alpha_{\sN}}{2\lambda}
  \right)
  v_{0}
  +
  \left(
    \frac{j_{0}}{2\lambda}
  \right)
  =
  0.
\end{equation}

\noindent
In the simple case in which we make $\lambda=0$, returning to the
free-field theory, we at once have that $\alpha_{\sN}=\alpha_{0}=\alpha$,
and the result reduces to

\begin{displaymath}
  j_{0}
  =
  \alpha\,
  v_{0},
\end{displaymath}

\noindent
which is the familiar result for the free theory.

\paragraph{Transversal Mass:} We are now in a position to discuss the
physical situation of the transversal renormalized mass in the general
case, in which $j_{0}$ and $v_{0}$ are not necessarily zero. This has to
be done separately in each phase of the model, and taking explicitly in
consideration whether or not there is a non-zero external source.

\subparagraph{Symmetric Phase:} In this case, if there is no external
source, then we have $v_{0}=0$ and therefore from
Equation~(\ref{ResultAlpha0}) the renormalized mass parameter is given by

\begin{equation}\label{Alpha0Result}
  \alpha_{0}
  =
  \alpha
  +
  \lambda
  \left[
    (\sN+1)
    \sigma_{0}^{2}
    +
    \sigma_{\sN}^{2}
  \right],
\end{equation}

\noindent
which is a positive quantity in this phase. On the other hand, if there is
an external source $j_{0}$, then there is also some value of $v_{0}$
associated to it, and therefore according to Equation~(\ref{ResultAlpha0})
the renormalized mass parameter changes to

\begin{displaymath}
  \alpha_{0}
  =
  \lambda
  v_{0}^{2}
  +
  \alpha
  +
  \lambda
  \left[
    (\sN+1)
    \sigma_{0}^{2}
    +
    \sigma_{\sN}^{2}
  \right].
\end{displaymath}

\noindent
This means that, given a point $(\alpha,\lambda)$ in the parameter space
of the model, the renormalized mass increases with $v_{0}$ and thus with
the external source. Note however that $\alpha_{0}$ does not depend
directly on the external source, but on $v_{0}$ instead. This indicates
that, in the case of localized external sources, the renormalized mass
should depend both on the external source and on the relative position
between the external source and the point of measurement of the mass.

\subparagraph{Broken-Symmetric Phase:} In this case, if there is no
external source, then instead of zero we have for $v_{0}$ the non-trivial
solution that we will denote here by $v_{0,{\rm SSB}}$, which according to
Equation~(\ref{ResultvSSB}) is given by

\begin{displaymath}
  \lambda
  v_{0,{\rm SSB}}^{2}
  =
  -
  \alpha
  -
  \lambda
  \left[
    (\sN-1)
    \sigma_{0}^{2}
    +
    3
    \sigma_{\sN}^{2}
  \right],
\end{displaymath}

\noindent
which is a positive quantity in this phase. Substituting this for the term
$\lambda v_{0}^{2}$ in Equation~(\ref{ResultAlpha0}) we get for the
transversal renormalized mass parameter

\begin{equation}\label{Alpha0SSB}
  \alpha_{0}
  =
  2\lambda
  \left(
    \sigma_{0}^{2}
    -
    \sigma_{\sN}^{2}
  \right).
\end{equation}

\noindent
Since $\sigma_{0}^{2}$ and $\sigma_{\sN}^{2}$ become identical in the
continuum limit, this seems to indicate that $\alpha_{0}$ tends to zero in
the limit and thus corresponds to zero mass $m_{0}$ in that limit.
However, $\alpha_{0}$ always goes to zero in the continuum limit, and the
fact that it does so is {\em not} enough to guarantee that $m_{0}$ is zero
in the limit. Therefore, further analysis of the limit in necessary, which
we will do later.

Going back to the case in which there is an external source $j_{0}$, we
see that $v_{0}$ will be somewhat larger that the solution $v_{0,{\rm
    SSB}}$. In this case we may add and subtract $\lambda v_{0,{\rm
    SSB}}^{2}$ in Equation~(\ref{ResultAlpha0}) and therefore write
$\alpha_{0}$ as

\begin{equation}\label{VarM1SSB}
  \alpha_{0}
  =
  \lambda
  \left(
    v_{0}^{2}
    -
    v_{0,{\rm SSB}}^{2}
  \right)
  +
  2\lambda
  \left(
    \sigma_{0}^{2}
    -
    \sigma_{\sN}^{2}
  \right),
\end{equation}

\noindent
showing once more that the mass increases with the variation of $v_{0}$
beyond its spontaneous symmetry-breaking value $v_{0,{\rm SSB}}$, and
hence that it increases with the introduction of the external source.
This represents the variation of $\alpha_{0}$ as a consequence of a
variation of $v_{0}$ beyond its spontaneous symmetry breaking value. In
terms of the mass $m_{0}$ this variation is not linear, but quadratic in
nature.

\paragraph{Longitudinal Mass:} Finally, we may now discuss the physical
situation of the longitudinal renormalized mass in the case in which
$v_{0}$ is not necessarily zero. This discussion proceeds in the same
lines as the previous one. Once more this has to be done separately in
each phase of the model, and taking in consideration whether or not there
is a non-zero external source.

\subparagraph{Symmetric Phase:} In this case, if there is no external
source, then we have $v_{0}=0$ and therefore from
Equation~(\ref{ResultAlphaN}) the renormalized mass parameter is given by

\begin{displaymath}
  \alpha_{\sN}
  =
  \alpha
  +
  \lambda
  \left[
    (\sN-1)
    \sigma_{0}^{2}
    +
    3
    \sigma_{\sN}^{2}
  \right],
\end{displaymath}

\noindent
which is a positive quantity in this phase. It is interesting to observe
that, since in the continuum limit $\sigma_{0}$ and $\sigma_{\sN}$ become
identical, for the purposes of that limit this equation is identical to
the corresponding result for $\alpha_{0}$, shown in
Equation~(\ref{Alpha0Result}), thus exhibiting the symmetry of the model
in this phase. On the other hand, if there is an external source, then
there is also some value of $v_{0}$ associated to it, and therefore
according to Equation~(\ref{ResultAlphaN}) the renormalized mass parameter
changes to

\begin{displaymath}
  \alpha_{\sN}
  =
  3\lambda
  v_{0}^{2}
  +
  \alpha
  +
  \lambda
  \left[
    (\sN-1)
    \sigma_{0}^{2}
    +
    3
    \sigma_{\sN}^{2}
  \right].
\end{displaymath}

\noindent
This means that, given a point $(\alpha,\lambda)$ in the parameter space
of the model, the renormalized mass increases with $v_{0}$ and thus with
the external source. This is now different from $\alpha_{0}$, since it
increases three times as fast with $v_{0}^{2}$. Note that once again
$\alpha_{\sN}$ does not depend directly on the external source, but on
$v_{0}$ instead.

\subparagraph{Broken-Symmetric Phase:} In this case, if there is no
external source, then instead of zero we have for $v_{0}$ the non-trivial
solution $v_{0,{\rm SSB}}$, which according to Equation~(\ref{ResultvSSB})
is given by

\begin{displaymath}
  \lambda
  v_{0,{\rm SSB}}^{2}
  =
  -
  \alpha
  -
  \lambda
  \left[
    (\sN-1)
    \sigma_{0}^{2}
    +
    3
    \sigma_{\sN}^{2}
  \right],
\end{displaymath}

\noindent
which is a positive quantity in this phase. Substituting this for the term
$\lambda v_{0}^{2}$ in Equation~(\ref{ResultAlphaN}) we get for the
longitudinal renormalized mass parameter

\begin{equation}\label{AlphaNResult}
  \alpha_{\sN}
  =
  -2
  \left\{
    \alpha
    +
    \lambda
    \left[
      (\sN-1)
      \sigma_{0}^{2}
      +
      3
      \sigma_{\sN}^{2}
    \right]
  \right\}.
\end{equation}

\noindent
This is a positive quantity in this phase, and in general corresponds to a
non-zero mass $m_{\sN}$. If, however, there is an external source, then
$v_{0}$ will be somewhat larger that the solution $v_{0,{\rm SSB}}$. In
this case we may add and subtract $3\lambda v_{0,{\rm SSB}}^{2}$ to
Equation~(\ref{ResultAlphaN}) and therefore write $\alpha_{\sN}$ as

\begin{displaymath}
  \alpha_{\sN}
  =
  3\lambda
  \left(
    v_{0}^{2}
    -
    v_{0,{\rm SSB}}^{2}
  \right)
  -2
  \left\{
    \alpha
    +
    \lambda
    \left[
      (\sN-1)
      \sigma_{0}^{2}
      +
      3
      \sigma_{\sN}^{2}
    \right]
  \right\},
\end{displaymath}

\noindent
showing once more that the mass increases with the variation of $v_{0}$
beyond its spontaneous symmetry-breaking value, and hence that it
increases with the introduction of the external source. If we denote the
value of $\alpha_{\sN}$ without the presence of the source by
$\alpha_{\sN,{\rm SSB}}$, we may write for the variation of $\alpha_{\sN}$
due to the external source

\begin{equation}\label{VarMNSSB}
  \alpha_{\sN}
  -
  \alpha_{\sN,{\rm SSB}}
  =
  3\lambda
  \left(
    v_{0}^{2}
    -
    v_{0,{\rm SSB}}^{2}
  \right).
\end{equation}

\noindent
Observe that once again this variation is three times larger than the
corresponding variation of $\alpha_{0}$.

\subsection{Continuum Limits}

First of all, it is necessary to say that, regardless of the spacetime
dimension and of the symmetry group which are chosen, the dimensionless
parameters $\alpha$ and $\lambda$ are the true free parameters of the
model. Besides the requirements of stability, there is no reason to limit
their range a priori. Limitations may arise, however, from the discussion
of physically meaningful observables, expressed as expectation values,
specially in the continuum limit. We start therefore with no more than the
stability conditions that $\lambda\geq 0$, and that $\alpha\geq 0$ if
$\lambda=0$, as the limitations for $\alpha$ and $\lambda$.

In the continuum limit, when we make $N\to\infty$ and $a\to 0$, most
dimensionless renormalized quantities we calculated here go to zero. In
order to recover the physically meaningful results in the limit, before we
take the limit we must rewrite these dimensionless quantities in terms of
the corresponding dimensionfull quantities, using the scalings listed in
Section~\ref{TheModel}, Equation~(\ref{Scalings}). Since in the continuum
limit $\sigma_{0}$ and $\sigma_{\sN}$ become identical, in all cases where
this is possible we will write the formulas in terms of $\sigma_{0}$ only,
producing in this way equations which are equivalent to the original ones
for the purposes of that limit.

Starting with the expectation value of the field, in the case in which
there is no external source $j_{0}$, in which case the limit must be taken
within the broken-symmetric phase of the model if we are to have the
possibility of a non-zero result, from Equation~(\ref{ResultvSSB}) we have

\noindent
\begin{eqnarray*}
  \left\langle\phi_{\sN}(x_{\mu})\right\rangle
  & = &
  V_{0}
  \\
  & = &
  \frac{v_{0}}{a^{(d-2)/2}}
  \\
  & = &
  \frac
  {\sqrt{-\left[\alpha+\lambda(\sN+2)\sigma_{0}^{2}\right]}}
  {\sqrt{\lambda}\,a^{(d-2)/2}}.
\end{eqnarray*}

\noindent
Since for $d\geq 3$ the denominator goes to zero in the continuum limit,
if the field $\phi_{\sN}(x_{\mu})$ is to have a finite expectation value,
then it is necessary that $v_{0}$ approach zero in the limit, which
forcefully takes us to points over the critical line, which is
characterized by $v_{0}=0$ and by the equation that states that the
quantity within the square root above is zero.

Since the critical line starts at the Gaussian point and extends to the
quadrant where $\lambda>0$ and $\alpha\leq 0$, it follows that all
possible continuum limits originating from the broken-symmetric phase must
go to points in the parameter plane where $\alpha\leq 0$, the case
$\alpha=0$ being the Gaussian point and corresponding to the Gaussian
sector of the model. In $d=4$, in particular, {\em all} possible
non-trivial continuum limits necessarily correspond to strictly negative
values of $\alpha$. A particular sequence of values of $(\alpha,\lambda)$
approaching the critical line defines both a path in the parameter space
of the model and a rate of progress along that path, leading to that
particular continuum limit, and is called a continuum limit {\em flow}. A
continuum limit is completely characterized by its flow, and is {\em not}
characterized completely just by a point $(\alpha,\lambda)$ in the
parameter plane.

Going back to the case in which we have an external source present, we may
now rewrite Equation~(\ref{ResultCubicEq}) in terms of the renormalized
dimensionfull quantities, thus obtaining

\begin{displaymath}
  a^{d-4}
  V_{0}^{3}
  -
  \left(
    \frac{m_{\sN}^{2}}{2\lambda}
  \right)
  V_{0}
  +
  \left(
    \frac{J_{0}}{2\lambda}
  \right)
  =
  0.
\end{displaymath}

\noindent
In the case $d=3$ we see that, if $\lambda$ is not zero, then the first
term dominates over the others, and therefore we conclude simply that
$V_{0}^{3}=0$. It follows that in this case there is no spontaneous
symmetry breaking and no effect of the external source over $V_{0}$ in the
continuum limit. If we wish to have any interesting structure in the model
in this case, we are forced to make $\lambda=0$ in the limit. If we do
that at the appropriate rate, there may be interesting continuum limits
sitting right over the Gaussian point. In the case $d\geq 5$, on the other
hand, we see that the first term vanishes, and we are left with
$J_{0}=m_{\sN}^{2}V_{0}$, which is characteristic of a free, or trivial
theory. In the case $d=4$ we get the equation

\begin{displaymath}
  V_{0}^{3}
  -
  \left(
    \frac{m_{\sN}^{2}}{2\lambda}
  \right)
  V_{0}
  +
  \left(
    \frac{J_{0}}{2\lambda}
  \right)
  =
  0.
\end{displaymath}

\noindent
It is interesting to calculate the discriminant $\Delta$ of this cubic
equation, which turns out to be

\begin{displaymath}
  \Delta
  =
  \frac{3^{3}}{2^{2}\lambda^{2}}
  \left(
    \sqrt{\frac{2}{3^{3}\lambda}}\,m_{\sN}^{3}+J_{0}
  \right)
  \left(
    \sqrt{\frac{2}{3^{3}\lambda}}\,m_{\sN}^{3}-J_{0}
  \right).
\end{displaymath}

\noindent
We can see now that the number of roots of the equation depends on the
value of $J_{0}$ in a simple way. If we have

\begin{displaymath}
  J_{0}
  <
  \sqrt{\frac{2}{3^{3}\lambda}}\,m_{\sN}^{3},
\end{displaymath}

\noindent
then $\Delta>0$ and therefore there are three distinct simple real
roots. If we have

\begin{displaymath}
  J_{0}
  =
  \sqrt{\frac{2}{3^{3}\lambda}}\,m_{\sN}^{3},
\end{displaymath}

\noindent
then $\Delta=0$ and the three roots merge into one triple real root.
Finally, if

\begin{displaymath}
  J_{0}
  >
  \sqrt{\frac{2}{3^{3}\lambda}}\,m_{\sN}^{3},
\end{displaymath}

\noindent
then $\Delta<0$ and there is a single real root, the other two having
non-zero imaginary parts. This supports the idea that as $J_{0}$ increases
along positive values, the left well of the potential becomes shallower
and eventually there is no possibility for the local distribution of the
field $\varphi_{\sN}$ to fit within it, even to form a meta-stable state.
One of the roots relates to the third extremum of the potential, the local
maximum between the two minima. It is clear that, when there is more than
one solution to the equation, only the largest solution corresponds to a
stable state and is therefore relevant in the context of the symmetry
breaking driven by a positive $J_{0}$.

The same analysis regarding critical behavior and the critical line is
valid for the renormalized masses. Considering first the limits from the
symmetric phase, with no external source $j_{0}$, we have
$\alpha_{0}=\alpha_{\sN}$ with $m_{0}^{2}=\alpha_{0}/a^{2}$, and therefore
using Equation~(\ref{Alpha0Result}) we have

\noindent
\begin{eqnarray*}
  \alpha_{0}
  & = &
  \alpha+\lambda(\sN+2)\sigma_{0}^{2}
  \Rightarrow
  \\
  m_{0}
  & = &
  \frac
  {\sqrt{\alpha+\lambda(\sN+2)\sigma_{0}^{2}}}
  {a}.
\end{eqnarray*}

\noindent
We can see that, regardless of how we take the limit, we will necessarily
have $m_{0}=m_{\sN}$ in this case. Observe that the numerator on the
right-hand side is the quantity which, according to the equation of the
critical line, is zero over that line, and hence approaches zero when
$(\alpha,\lambda)$ tends to a point on the critical line. Once more we see
that, if we are to have a finite value for $m_{0}$, we must approach the
critical line on the continuum limit, in such a way that the quantity
$\alpha+\lambda(\sN+2)\sigma_{0}^{2}$ approaches zero as $a^{2}$ or
faster. If the approach is such that the quantity in the numerator behaves
exactly as $a^{2}$, then we have a finite and non-zero value of $m_{0}$.
If the approach is faster than that, then we will have $m_{0}=0$. On the
other hand, if it is too slow, then we may end up with an infinite $m_{0}$
in the limit.

The same type of mechanism works for limits from the broken-symmetric
phase, except that in that case we will always have $m_{0}=0$ in the
limit, as we will now demonstrate. As we saw before in
Equation~(\ref{Alpha0SSB}), we have for $\alpha_{0}$

\begin{displaymath}
  \alpha_{0}
  =
  2\lambda
  \left(
    \sigma_{0}^{2}
    -
    \sigma_{\sN}^{2}
  \right),
\end{displaymath}

\noindent
which indeed goes to zero in the limit. However, the analysis of the limit
is not so simple, due to the fact that on finite lattices $\alpha_{0}$
appears in the right-hand side of the equation as well. If we write it
explicitly, using Equations~(\ref{Sigma0Form}) and~(\ref{SigmaNForm}) of
Appendix~\ref{TabofInts}, we get an equation involving $\alpha_{0}$ and
$\alpha_{\sN}$,

\noindent
\begin{eqnarray*}
  \alpha_{0}
  & = &
  2\lambda\,
  \frac{1}{N^{d}}
  \sum_{k_{\mu}}^{N^{d}}
  \left[
    \frac{1}{\rho^{2}(k_{\mu})+\alpha_{0}}
    -
    \frac{1}{\rho^{2}(k_{\mu})+\alpha_{\sN}}
  \right]
  \\
  & = &
  2\lambda
  \left(\alpha_{\sN}-\alpha_{0}\right)
  \frac{1}{N^{d}}
  \sum_{k_{\mu}}^{N^{d}}
  \frac{1}
  {
    \left[\rho^{2}(k_{\mu})+\alpha_{0}\right]
    \left[\rho^{2}(k_{\mu})+\alpha_{\sN}\right]
  }.
\end{eqnarray*}

\noindent
Now, if $\alpha_{\sN}=\alpha_{0}$, which implies that $m_{\sN}=m_{0}$,
then the right-hand side is zero, and therefore so is $\alpha_{0}$. This
in turn implies that $m_{0}=0$, as expected. This is in fact one
possibility, we may indeed have both $m_{0}$ and $m_{\sN}$ zero in the
limit. If, on the other hand, we have $\alpha_{\sN}\neq\alpha_{0}$, then
we may write the equation as

\begin{displaymath}
  \frac{m_{0}^{2}}{m_{\sN}^{2}-m_{0}^{2}}
  =
  2\lambda
  \frac{1}{N^{d}}
  \sum_{k_{\mu}}^{N^{d}}
  \frac{1}
  {
    \left[\rho^{2}(k_{\mu})+\alpha_{0}\right]
    \left[\rho^{2}(k_{\mu})+\alpha_{\sN}\right]
  },
\end{displaymath}

\noindent
where we wrote the left-hand side in terms of dimensionfull quantities.
Obviously, because both $\lambda$ and the sum are necessarily positive
quantities, it is impossible to have $m_{\sN}<m_{0}$. Here we see that, if
we have both $m_{0}$ and $m_{\sN}>m_{0}$ different from zero in the limit,
then the left-hand side has a non-zero limit and therefore the normalized
sum on the right-hand side must be non-zero in the limit.

However, one can check numerically that, for $d\geq 4$ and in the type of
continuum limit that we consider here, the normalized sum does indeed go
to zero in the limit. This implies that in these dimensions, which include
$d=4$, we cannot have both $m_{0}$ and $m_{\sN}$ different from zero in
the limit. Since $m_{\sN}>m_{0}$, this implies that we must always have
$m_{0}=0$ in the limit. What we have here, as one should expect, are the
Goldstone bosons brought about by the process of spontaneous symmetry
breaking.

For the longitudinal mass parameter $m_{\sN}$ we have, using
Equation~(\ref{AlphaNResult}),

\noindent
\begin{eqnarray*}
  \alpha_{\sN}
  & = &
  -2
  \left[\alpha+\lambda(\sN+2)\sigma_{0}^{2}\right]
  \Rightarrow
  \\
  m_{\sN}
  & = &
  \frac
  {\sqrt{-2\left[\alpha+\lambda(\sN+2)\sigma_{0}^{2}\right]}}
  {a},
\end{eqnarray*}

\noindent
so that exactly the same argument that was used for $\alpha_{0}$ in the
symmetric phase applies. We see therefore that the need to approach the
critical line when one takes continuum limits in this model is a rather
general characteristic of the model. This makes the critical line the
locus of {\em all} physically possible continuum limits of the model.
This means that making $\alpha<0$ is not a choice that we have, since it
is {\em forced} upon us by the need to obtain physically meaningful
continuum limits.

Let us now discuss the continuum limits of the transversal renormalized
mass in the presence of an external source. We have the result in
Equation~(\ref{ResultAlpha0}), valid in either phase,

\begin{displaymath}
  \alpha_{0}(j_{0})
  =
  \lambda
  v_{0}^{2}
  +
  \left[
    \alpha
    +
    \lambda
    (\sN+2)
    \sigma_{0}^{2}
  \right].
\end{displaymath}

\noindent
Observe that this equation implies that it is still necessary to approach
the critical line in the continuum limit, and in the same ways as before.
In the symmetric phase, if $\alpha_{0}(0)$ is the corresponding result in
the absence of external sources, which corresponds to $v_{0}=0$, we may
write

\begin{displaymath}
  \alpha_{0}(j_{0})
  =
  \alpha_{0}(0)
  +
  \lambda
  v_{0}^{2}.
\end{displaymath}

\noindent
Rewriting all quantities in terms of the corresponding dimensionfull ones
we have

\begin{displaymath}
  m_{0}^{2}(J_{0})
  =
  m_{0}^{2}(0)
  +
  \lambda
  a^{d-4}
  V_{0}^{2}.
\end{displaymath}

\noindent
We see that for $d=3$ we are forced to make $\lambda\to 0$ in the limit.
In the case $d=4$ no additional constraints on $\lambda$ arise, and we get
the relation

\begin{displaymath}
  m_{0}(J_{0})
  =
  \sqrt
  {
    m_{0}^{2}(0)
    +
    \lambda
    V_{0}^{2}
  },
\end{displaymath}

\noindent
describing indirectly how $m_{0}(J_{0})$ increases with $J_{0}$ through
the variation of $V_{0}$. In the case $d=5$ the term containing $\lambda$
vanishes in the limit, and we get simply that $m_{0}(J_{0})=m_{0}(0)$,
meaning that in this case $m_{0}(J_{0})$ does not really depend on $J_{0}$
in the continuum limit.

In the broken-symmetric phase we may start with Equation~(\ref{VarM1SSB})
for the transversal mass parameter. If we recall that we have already
shown that in this phase we must have $m_{0}(0)=0$ in the limit, we may
make $\sigma_{\sN}=\sigma_{0}$ in this formula and thus obtain

\begin{displaymath}
  \alpha_{0}
  =
  \lambda
  \left(
    v_{0}^{2}
    -
    v_{0,{\rm SSB}}^{2}
  \right).
\end{displaymath}

\noindent
In terms of dimensionfull quantities we have therefore

\begin{displaymath}
  m_{0}^{2}(J_{0})
  =
  \lambda
  a^{d-4}
  \left(
    V_{0}^{2}
    -
    V_{0,{\rm SSB}}^{2}
  \right),
\end{displaymath}

\noindent
which gives us back $m_{0}(0)=0$ in the absence of external sources. Not
much changes in the discussion of the various possible dimensions. We may
restrict our comments to the case $d=4$, in which we get a fairly simple
relation giving $m_{0}(J_{0})$ in the presence of the external source,

\begin{displaymath}
  m_{0}(J_{0})
  =
  \sqrt{\lambda}
  \sqrt
  {
    V_{0}^{2}
    -
    V_{0,{\rm SSB}}^{2}
  }.
\end{displaymath}

\noindent
The same analysis can be made for the longitudinal mass in the presence of
an external source. In this case we have the result in
Equation~(\ref{ResultAlphaN}), valid in either phase,

\begin{displaymath}
  \alpha_{\sN}(j_{0})
  =
  3\lambda
  v_{0}^{2}
  +
  \left[
    \alpha
    +
    \lambda
    (\sN+2)
    \sigma_{0}^{2}
  \right].
\end{displaymath}

\noindent
The necessity to approach the critical line remains in force here. In the
symmetric phase, if $\alpha_{\sN}(0)$ is the corresponding result in the
absence of external sources, which corresponds to $v_{0}=0$, we may write

\begin{displaymath}
  \alpha_{\sN}(j_{0})
  =
  \alpha_{\sN}(0)
  +
  3\lambda
  v_{0}^{2}.
\end{displaymath}

\noindent
Rewriting all quantities in terms of the corresponding dimensionfull ones
we have

\begin{displaymath}
  m_{\sN}^{2}(J_{0})
  =
  m_{\sN}^{2}(0)
  +
  3\lambda
  a^{d-4}
  V_{0}^{2}.
\end{displaymath}

\noindent
Once more we see that for $d=3$ we are forced to make $\lambda\to 0$ in
the limit. In the case $d=4$ we get simply the relation

\begin{displaymath}
  m_{\sN}(J_{0})
  =
  \sqrt
  {
    m_{\sN}^{2}(0)
    +
    3\lambda
    V_{0}^{2}
  },
\end{displaymath}

\noindent
describing indirectly how $m_{\sN}(J_{0})$ increases with $J_{0}$ through
the variation of $V_{0}$. In the case $d=5$ the term containing $\lambda$
vanishes in the limit, and we get simply that $m_{\sN}(J_{0})=m_{\sN}(0)$,
meaning that in this case $m_{\sN}(J_{0})$ also does not depend on $J_{0}$
in the continuum limit.

In the broken-symmetric phase we may start with Equation~(\ref{VarMNSSB})
for the longitudinal mass parameter

\begin{displaymath}
  \alpha_{\sN}
  -
  \alpha_{\sN,{\rm SSB}}
  =
  3\lambda
  \left(
    v_{0}^{2}
    -
    v_{0,{\rm SSB}}^{2}
  \right).
\end{displaymath}

\noindent
where $\alpha_{\sN,{\rm SSB}}$ is the value of the parameter in the
absence of external sources. In terms of dimensionfull quantities we have
therefore

\begin{displaymath}
  m_{\sN}^{2}(J_{0})
  -
  m_{\sN}^{2}(0)
  =
  3\lambda
  a^{d-4}
  \left(
    V_{0}^{2}
    -
    V_{0,{\rm SSB}}^{2}
  \right),
\end{displaymath}

\noindent
Once again not much changes in the discussion of the various possible
dimensions. In the case $d=4$ we get

\begin{displaymath}
  m_{\sN}(J_{0})
  =
  \sqrt
  {
    m_{\sN}^{2}(0)
    +
    3\lambda
    \left(
      V_{0}^{2}
      -
      V_{0,{\rm SSB}}^{2}
    \right)
  }.
\end{displaymath}

\noindent
It is interesting to note that, both for the transversal and longitudinal
masses, the dependence of the renormalized masses on the external source
$J_{0}$ seems to be a peculiar feature of the case $d=4$, which is absent
for $d\geq 5$.

\section{Some Consequences}

The calculations performed in this paper have a few rather interesting
consequences, and some relevant conclusions can be drawn from them. In
this section we discuss some of these consequences.

\subsection{Triviality Tests}

The fact that the renormalized masses depend on the external sources, as
we saw above, has important consequences for the design of computer
simulations targeted at probing the triviality issue. One way to do this
is to perform simulations on finite lattices in which one tries to measure
the relation between the external sources and the expectation value of the
field. The argument is based on the fact that on symmetry grounds it is
reasonable to expect that the model has an effective action with the
general form

\noindent
\begin{eqnarray*}
  \Gamma_{N}[\vec{\varphi}_{\rm c}]
  & = &
  \sum_{n_{\mu}}^{N^{d}}
  \left\{
    \rule{0em}{4ex}
    \frac{1}{2}
    \sum_{\nu}^{d}
    \left[
      \Delta_{\nu}\vec{\varphi}_{\rm c}(n_{\mu})
      \cdot
      \Delta_{\nu}\vec{\varphi}_{\rm c}(n_{\mu})
    \right]
    +
  \right.
  \\
  &   &
  \hspace{2.0em}
  \left.
    \rule{0em}{4ex}
    +
    \frac{\alpha_{0}}{2}
    \left[\vec{\varphi}_{\rm c}(n_{\mu})\cdot\vec{\varphi}_{\rm c}(n_{\mu})\right]
    +
    \frac{\alpha_{\sN}-\alpha_{0}}{2}\,
    \varphi_{\sN,{\rm c}}^{2}(n_{\mu})
    +
  \right.
  \\
  &   &
  \hspace{2.0em}
  \left.
    \rule{0em}{4ex}
    +
    \frac{\lambda_{R}}{4}
    \left[\vec{\varphi}_{\rm c}(n_{\mu})\cdot\vec{\varphi}_{\rm c}(n_{\mu})\right]^{2}
    -
    j_{0}\varphi_{\sN,{\rm c}}(n_{\mu})
  \right\},
\end{eqnarray*}

\noindent
where $\vec{\varphi}_{\rm c}(n_{\mu})$ is the classical field, which is
just another name for the expectation value of the field, given an
arbitrary external source $j_{0}$. In this expression $\alpha_{0}$ and
$\alpha_{\sN}$ are the renormalized masses, and $\lambda_{R}$ is the
renormalized coupling constant. It is possible that additional terms may
appear in $\Gamma_{N}[\vec{\varphi}_{\rm c}]$, but terms containing
derivatives are not relevant for the argument that follows, and terms with
higher powers can be easily included in the analysis if need be.

If one considers only homogeneous external sources $j_{0}$, then it is
clear that the classical field must also be a constant, $\varphi_{\sN,{\rm
    c}}(n_{\mu})=v_{0}$, and hence all terms containing derivatives
vanish. We are left with only the part of the effective action that
contains the effective potential. In addition to this, since the external
source is in the direction of $\varphi_{\sN,{\rm c}}(n_{\mu})$ in the
internal $SO(\sN)$ space, it is clear that the expectation values of all
the other field components are zero, so that we are left with

\begin{displaymath}
  \Gamma_{N}[\vec{\varphi}_{\rm c}]
  =
  \sum_{n_{\mu}}^{N^{d}}
  \left[
    \frac{\alpha_{\sN}}{2}\,
    v_{0}^{2}
    +
    \frac{\lambda_{R}}{4}\,
    v_{0}^{4}
    -
    j_{0}v_{0}
  \right],
\end{displaymath}

\noindent
where we now wrote $v_{0}$ for the expectation value of the field. Since
the behavior of $\varphi_{\sN,{\rm c}}(n_{\mu})$ is ruled by the minimum
of this action in the classical or long-wavelength limit, which is
consistent with the use of a homogeneous external source, we may now
differentiate with respect to the classical field $v_{0}$, and equate the
result to zero, thus obtaining

\begin{displaymath}
  j_{0}
  =
  \alpha_{\sN}\,
  v_{0}
  +
  \lambda_{R}\,
  v_{0}^{3}.
\end{displaymath}

\noindent
By measuring $v_{0}$ as a function of $j_{0}$ one may then determine from
this equation the coefficients $\alpha_{\sN}$ and $\lambda_{R}$, and thus
probe into the triviality of the model. For $d\geq 4$ triviality would
result if it can be shown that

\begin{displaymath}
  \lim_{N\to\infty}\lambda_{R}
  =
  0.
\end{displaymath}

\noindent
In other words, a linear result for the relation between $j_{0}$ and
$v_{0}$, with the renormalized mass parameter as the coefficient,
indicates triviality. Any deviations from linearity imply the existence of
interactions, either on finite lattices or in the continuum limit. This
technique avoids the necessity for the direct measurement of the
four-point function, which is generally much more difficult to do
numerically than to measure the one-point and two-point functions.

However, we have shown here that $\alpha_{\sN}$ itself depends on
$j_{0}$. Therefore, even if $\lambda_{R}$ is in fact zero on finite
lattices this relation will not result linear if the simulations are
performed by varying $j_{0}$ at fixed values of parameters $\alpha$ and
$\lambda$. It is therefore necessary to adjust these parameters, as one
varies $j_{0}$, in order to keep $\alpha_{\sN}$ constant. For this purpose
the value of $\alpha_{\sN}$ can be obtained independently via the
measurement of the propagator of the field component
$\varphi_{\sN}(n_{\mu})$, of course. At the end of the day, its value can
be confirmed by the value resulting for the linear coefficient from a
polynomial fit to the relation between $j_{0}$ and $v_{0}$.

Given a certain chosen value for $\alpha_{\sN}$, for each value of $j_{0}$
one must search the parameter plane of the model looking for a point where
$\alpha_{\sN}$ has that value, and only then measure $v_{0}$. This can
become a computationally expensive search. This can be done by keeping
$\lambda$ constant and varying $\alpha$, thus traversing horizontal lines
on the parameter plane, or by keeping $\alpha$ constant and varying
$\lambda$, thus traversing vertical lines. Given the structure of the
phase transition and of the critical line, one attractive alternative is
to keep $\alpha^{2}+\lambda^{2}$ constant and vary the angle $\theta$
around the position of the critical line, where

\begin{displaymath}
  \frac{\lambda}{-\alpha}
  =
  \tan(\theta),
\end{displaymath}

\noindent
not forgeting that in general $\alpha$ will be negative. In any case, the
formula giving $\alpha_{\sN}$ in terms of $\alpha$, $\lambda$, $v_{0}$ and
$j_{0}$ that we derived here, shown in Equation~(\ref{ResultAlphaN}), may
serve to provide at least a good initial guess for this costly search in
the parameter plane.

\subsection{Standard Model}

The four-component $SO(4)$ model has an important application in the
Standard Model of high-energy elementary particles. The field component
$\varphi_{\sN}(n_{\mu})$ corresponds in this case to the Higgs field. In
this application the continuum limit must be taken from the
broken-symmetric phase, for it is essential that we have, in the limit, a
non-zero $V_{0}$ due to spontaneous symmetry breaking.

It is certainly possible to take limits from the broken-symmetric phase to
the critical line in such a way that either $V_{0}$ or $m_{\sN}$ has a
finite and non-zero limit. It is not so obvious, but true in $d=4$, that
one can take limits in which {\em both} are simultaneously finite and
non-zero. In fact, the calculations imply that in this case there is a
definite relation between $V_{0}$ and $m_{\sN}$.

If we recall our results for $v_{0}$ and $\alpha_{\sN}$ in the
broken-symmetric phase (Equations~(\ref{ResultvSSB})
and~(\ref{AlphaNResult})), without external sources, we have

\noindent
\begin{eqnarray*}
  \lambda
  v_{0}^{2}
  & = &
  -
  \left[
    \alpha
    +
    \lambda
    (\sN+2)
    \sigma_{0}^{2}
  \right],
  \\
  \alpha_{\sN}
  & = &
  -2
  \left[
    \alpha
    +
    \lambda
    (\sN+2)
    \sigma_{0}^{2}
  \right].
\end{eqnarray*}

\noindent
It immediately follows that we have the following result relating $v_{0}$
and $\alpha_{\sN}$,

\begin{displaymath}
  2\lambda
  v_{0}^{2}
  =
  \alpha_{\sN}.
\end{displaymath}

\noindent
Writing this in terms of dimensionfull quantities we get

\begin{displaymath}
  2\lambda
  a^{d-4}
  V_{0}^{2}
  =
  m_{\sN}^{2}.
\end{displaymath}

\noindent
Of course the important dimension here is $d=4$, but let us comment on the
other cases anyway. In $d=3$ we are forced once again to make $\lambda\to
0$, which takes us to the Gaussian point, and if we do this at the
appropriate pace, we then simply get $2\Lambda V_{0}^{2}=m_{\sN}^{2}$. In
$d=5$ we conclude that, so long as $\lambda$ and $V_{0}$ are finite, we
must have $m_{\sN}=0$. If we insist on a finite and non-zero $m_{\sN}$,
then $V_{0}$ must diverge to infinity. So in this case we cannot take a
limit in such a way that both $V_{0}$ and $m_{\sN}$ remain finite and
non-zero.

However, in $d=4$, and only in $d=4$, we get a definite relation between
$V_{0}$ and $m_{\sN}$, involving only the dimensionless parameters of the
model, and valid for all allowed values of these parameters within the
broken-symmetric phase, given by

\begin{displaymath}
  \frac{V_{0}}{m_{\sN}}
  =
  \frac{1}{\sqrt{2\lambda}}.
\end{displaymath}

\noindent
Since the values of $V_{0}$ and $m_{\sN}$ are known experimentally, namely
$V_{0}\approx 246$~Gev and $m_{\sN}\approx 126$~Gev, we immediately get a
result for $\lambda$,

\begin{displaymath}
  \lambda
  \approx
  0.131.
\end{displaymath}

\noindent
Given this result, we can find $\alpha$ as well. All we have to do is to
use the equation of the critical line, given in
Equation~(\ref{CriticalLine}),

\begin{displaymath}
  \alpha
  +
  \lambda
  (\sN+2)
  \sigma_{0}^{2}
  =
  0,
\end{displaymath}

\noindent
with $\sN=4$ and our best numerical evaluation of $\sigma_{0}^{2}$ for
$d=4$, which is $\sigma_{0}^{2}\approx 0.15493$, and we get

\begin{displaymath}
  \alpha
  \approx
  -0.122.
\end{displaymath}

\noindent
Conceptually, this is a rather remarkable result. Please observe that we
are not using the experimental data to make statements about expectation
values, but instead to determine the values of bare dimensionless
parameters within the mathematical structure of the model. We are able,
using the experimental data, to pinpoint the location in the parameter
space of the model, along the critical line, where it must be located if
it is applicable to the real world,

\begin{displaymath}
  (\alpha,\lambda)
  \approx
  (-0.122,0.131).
\end{displaymath}

\begin{figure}[ht]
  \centering
  \fbox{
    \epsfig{file=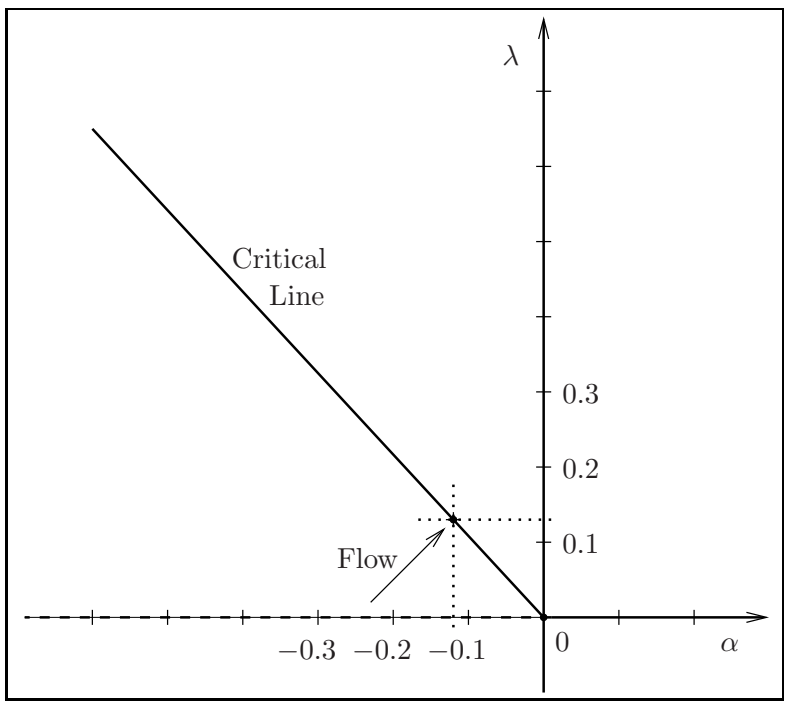,scale=1.0,angle=0}
  }
  \caption{Critical diagram of the model in $d=4$, for $\sN=4$, with the
    Standard Model continuum limit point
    $(\alpha,\lambda)\approx(-0.122,0.131)$ singled out, showing the path
    of a possible continuum limit flow.}
  \label{CritFig}
\end{figure}

\noindent
This is a point at a distance of approximately $0.179$ from the Gaussian
point, along the critical line, which makes an angle of approximately
$47.0$ degrees with the negative $\alpha$ semi-axis. The situation in the
parameter-plane of the model is depicted in Figure~\ref{CritFig}, which is
drawn approximately to scale.

One may wonder how accurate this result may be. In the Standard Model
there are electroweak charges associated to $\vec{\varphi}(n_{\mu})$,
which are being ignored here. It is of course possible that these other
interactions might change the expectation value and the renormalized mass
of the Higgs field. However, after the symmetry is broken and the three
Goldstone bosons $\varphi_{i}(n_{\mu})$, $i=1,2,3$ are absorbed by the
three massive vector bosons, the single remaining scalar field which is
the Higgs has no electromagnetic charge, and undergoes only weak
interactions, if any. Therefore it is reasonable to think that whatever
corrections there may be to the result above are probably quite small. By
comparison to possible weak perturbative corrections, the results
presented here have a rather brutal character, since they handle correctly
the non-perturbative phenomenon of spontaneous symmetry breaking, at the
quantum level, flipping the sign of $\alpha$ to negative values in that
process.

\subsection{Hints on Triviality}

Since we have results for both $v_{0}$ and $m_{\sN}$ as functions of
$j_{0}$, it is conceivable that these results can give us some hints as to
the question of triviality. As we shall see, trying to do this does in
fact provide some rather crude hints, but most of all it puts in evidence
the limitations of the calculational technique.

Our Gaussian-Perturbative result for the relation between $j_{0}$ and
$v_{0}$ in either phase of the model, as shown in
Equation~(\ref{ResultCL}), may be written for the purposes of the
continuum limit as

\begin{displaymath}
  j_{0}
  =
  v_{0}
  \left\{
    \lambda
    v_{0}^{2}
    +
    \left[
      \alpha
      +
      \lambda
      (\sN+2)
      \sigma_{0}^{2}
    \right]
  \right\}.
\end{displaymath}

\noindent
On the other hand, our result for the renormalized mass parameter
$\alpha_{\sN}$, also valid in either phase of the model, in the presence
of the external source, as shown in Equation~(\ref{ResultAlphaN}), may be
written for the purposes of the continuum limit as

\begin{displaymath}
  \alpha_{\sN}
  =
  3\lambda
  v_{0}^{2}
  +
  \left[
    \alpha
    +
    \lambda
    (\sN+2)
    \sigma_{0}^{2}
  \right].
\end{displaymath}

\noindent
It follows that we may combine these two results, and write a relation
involving the renormalized quantities $v_{0}$ and $\alpha_{\sN}$,

\begin{displaymath}
  j_{0}
  =
  v_{0}
  \left(
    \alpha_{\sN}
    -
    2\lambda
    v_{0}^{2}
  \right).
\end{displaymath}

\noindent
This relation is valid for all $d\geq 3$, for all $\sN\geq 1$, and all
explicit references to $\alpha$ are gone. In the continuum limit both
sides of this equation approach zero. In order to analyze the limit, it is
necessary to rewrite everything in terms of the corresponding finite and
possibly non-zero dimensionfull quantities. Doing this with the use of the
scalings given in Section~\ref{TheModel}, Equation~(\ref{Scalings}), we
get

\begin{displaymath}
  J_{0}
  =
  V_{0}
  \left(
    m_{\sN}^{2}
    -
    2\lambda\,
    a^{d-4}
    V_{0}^{2}
  \right).
\end{displaymath}

\noindent
This behaves differently in each dimension $d$. We will analyze separately
the cases $d=3$, $d=4$ and $d\geq 5$.

\paragraph{Case $d=3$:} The result becomes inconsistent unless we make
$\lambda\to 0$ in the continuum limit, which takes us to the Gaussian
point. If we do that sufficiently fast then we get
$J_{0}=m_{\sN}^{2}V_{0}$, a result consistent with a linear theory.
Otherwise, if we take $\lambda$ to zero at the appropriate rate, we get

\begin{displaymath}
  J_{0}
  =
  V_{0}
  \left(
    m_{\sN}^{2}
    -
    2\Lambda\,
    V_{0}^{2}
  \right).
\end{displaymath}

\noindent
Since the sign of the second term is reversed, this result does not seem
to make much sense, even if we consider that the $\Lambda$ that appears
there is a bare parameter, a parameter characterizing a continuum-limit
flow, in fact, and {\em not} the renormalized coupling constant.

Observe however that this is not necessarily a free theory, even at the
Gaussian point, where one would expect that $\lambda_{R}=0$. This is so
because in $d=3$ we have $\Lambda_{R}=\lambda_{R}/a$, and therefore we may
have $\Lambda_{R}\neq 0$ even if $\lambda_{R}\to 0$ as we take the limit
and make $a\to 0$. In other words, in $d=3$ we may have interacting limits
of this type sitting right on top of the Gaussian point.

The complete failure of the approximation away from the Gaussian point
suggests that in that case the distribution of the $d=3$ model may not be
sufficiently close to a Gaussian distribution to allow a Gaussian
approximation to work, and hence its expectations values cannot be well
represented by the Gaussian measure of $S_{0}[\vec{\varphi}']$.

\paragraph{Case $d=4$:} The result is perfectly regular, and is simply
given by

\begin{displaymath}
  J_{0}
  =
  V_{0}
  \left(
    m_{\sN}^{2}
    -
    2\lambda\,
    V_{0}^{2}
  \right).
\end{displaymath}

\noindent
Not much can be concluded in this case, though. Of course we must not
forget that both $V_{0}$ and $m_{\sN}$ are functions of $\lambda$ and
$J_{0}$, in such a way that the right-hand side of this equation remains
positive for positive $J_{0}$.

Although this equation has the general form expected for an interacting
theory, it is crucial to note that the sign of the second term is
reversed. Since we must have $\lambda>0$, this term is necessarily
negative. This is not really all that surprising, for one must not forget
that it is not to be expected that this calculational technique can
produce predictions about $\lambda_{R}$, which is a parameter related to
the fourth moment of the distribution, that is of course absent from a
Gaussian approximation. The reversed sign, that seems to indicate that
increasing $\lambda$ works in the way opposite to what one would expect,
appears there as a warning about this limitation. Of course, interpreting
this term as a prediction for $\lambda_{R}$ would be absurd, since it
would imply that the renormalized coupling constant is negative, and thus
would correspond to an unstable renormalized model.

\paragraph{Case $d\geq 5$:} The result is not only perfectly regular, but
the second term in the right-hand side vanishes in the limit, so long as
$\lambda$ is kept finite, and one is left with the simple result
$J_{0}=m_{\sN}^{2}V_{0}$, which is consistent with a trivial theory. In
this case this result is valid for any finite value of $\lambda$. This is
consistent with the triviality of the model for $d\geq 5$, which seems to
be a fairly well-established fact.

It is interesting to observe that it is possible to define a version of
this model in which the limit $\lambda\to\infty$ is taken. It is possible
to show, with all mathematical rigor, and for any $d$ and any $\sN$, that
the limit of the $SO(\sN)$ polynomial model we have here, in which one
makes $\lambda\to\infty$ and $\alpha\to-\infty$ in such a way that
$\beta=-\alpha/\lambda$ is kept finite, is in fact the $SO(\sN)$
non-linear Sigma Model with coupling constant $\beta$. It is therefore
possible that the $SO(\sN)$ non-linear Sigma Models in $d=5$ or more may
still have some interesting continuum limits.

\section{Conclusions}

It was already known, for some time, that the approximation scheme that we
name here the Gaussian-Perturbative approximation gives good results for
the $SO(\sN)$-symmetric $\lambda\phi^{4}$ model in $d=4$, regarding its
critical behavior~\cite{pertheory}. It is interesting to speculate that
the triviality of the model in $d=4$, which is fairly well established
numerically but still lacks rigorous proof, is somehow behind the fact
that this approximation works as well as it does in that case. This is so
because a trivial model would have a Gaussian effective action, which
would allow for a good approximation of its expectation values by a
Gaussian measure, which is what we do in the Gaussian-Perturbative
approximation.

In this work we extended that technique to the same model in the presence
of an external source. This resulted in specific predictions for the
values of the expectation value of the field and for the renormalized
masses as functions of the external source. Such predictions could
motivate future numerical studies with the objective of evaluating their
worth by comparing them to the results of appropriate stochastic
simulations. In particular, the fact that the renormalized masses do
depend on the external sources through the expectation value of the field
is important for the very design of some such numerical simulations.

The simulations done in the past to test this technique used what we named
back-rotation simulations, which introduce some additional uncertainties
into the whole analysis. This was done because neither in the analytical
calculations nor in the numerical simulations we were capable at that time
to deal appropriately with the external sources. It is now possible to
perform simulations in the presence external sources, without the use of
the back-rotation idea. With such simulations and the results presented in
this paper, it should be possible to do much better comparison of the
numerical and analytical results.

We also pointed out a simple and interesting consequence of the results
regarding the application of the $\lambda\phi^{4}$ model in the Standard
Model of particle physics. The results allow us to determine the critical
point $(\alpha,\lambda)$ in the parameter plane of the model that should
correspond to the continuum-limit flows leading to the Standard Model.
This is a rather unique situation, in which actual experimental data is
used to determine the values of bare, dimensionless parameters within the
mathematical structure of the $\lambda\phi^{4}$ model.

Although this result in itself may be no more than a curiosity, it would
be interesting to determine whether or not this technique and its results
could not find a more widespread application for the computation of
physical predictions from the Standard Model. This, combined with the use
of the very probable fact that the renormalized coupling constant
$\lambda_{R}$ is in fact exactly zero in the continuum limits of this
model, could very well result in the extraction of new and interesting
insights from the Standard Model.

As an example, we may point out that the attribution of a negative value
to the bare parameter $\alpha$ is not really a matter of choice, as is
implied in the usual treatment of the Standard Model. It is in fact a very
{\em strict requirement} for the existence of physically meaningful
continuum limits of the model, as we have shown in this paper. There are
in fact {\em no} continuum limits, in which the fundamental action is not
Gaussian and the model has finite renormalized masses, such that
$\alpha>0$ in the limit.

Is in important to point out that the results presented here for critical
behavior and symmetry breaking within the $\lambda\phi^{4}$ model are
quite independent of the renormalized coupling constant $\lambda_{R}$. In
particular, they are quite independent of whether or not $\lambda_{R}$ is
zero in the continuum limit. In other words, the probable triviality of
the model in the continuum limit does not disturb the mechanism of phase
transition and symmetry breaking, and hence would not void the Higgs
mechanism.

\appendix

\renewcommand{\theequation}{\Alph{section}.\arabic{equation}}

\section{Calculation of Expectation Values}\label{ExpecVals}

\setcounter{equation}{0}

In this section we calculate in detail the expectation values which are
needed for the evaluation of the observables discussed in this paper.
These are all expectation values in the measure of the Gaussian action
given in Equation~(\ref{ResultS0}),

\noindent
\begin{eqnarray*}
  S_{0}[\vec{\varphi}']
  & = &
  \sum_{n_{\mu}}^{N^{d}}
  \left\{
    \rule{0em}{4ex}
    \frac{1}{2}
    \sum_{\nu}^{d}
    \left[
      \Delta_{\nu}\vec{\varphi}'(n_{\mu})
      \cdot
      \Delta_{\nu}\vec{\varphi}'(n_{\mu})
    \right]
    +
  \right.
  \\
  &   &
  \hspace{2.0em}
  \left.
    \rule{0em}{4ex}
    +
    \frac{\alpha_{0}}{2}
    \left[\vec{\varphi}'(n_{\mu})\cdot\vec{\varphi}'(n_{\mu})\right]
    +
    \frac{\alpha_{\sN}-\alpha_{0}}{2}\,
    \varphi_{\sN}'^{2}(n_{\mu})
  \right\},
\end{eqnarray*}

\noindent
which is even on the fields. They will all involve the non-Gaussian or
``interacting'' part of the action, which is given in
Equation~(\ref{ResultSV}),

\noindent
\begin{eqnarray*}
  S_{V}[\vec{\varphi}']
  & = &
  \sum_{n_{\mu}}^{N^{d}}
  \left\{
    \rule{0em}{3ex}
    v_{0}
    \left[\alpha+\lambda v_{0}^{2}\right]
    \varphi_{\sN}'(n_{\mu})
    -
    j_{0}\varphi_{\sN}'(n_{\mu})
    +
  \right.
  \\
  &   &
  \hspace{2.23em}
  \left.
    +
    \frac{\alpha-\alpha_{0}+\lambda v_{0}^{2}}{2}
    \left[\vec{\varphi}'(n_{\mu})\cdot\vec{\varphi}'(n_{\mu})\right]
    +
    \frac{\alpha_{0}-\alpha_{\sN}+2\lambda v_{0}^{2}}{2}\,
    \varphi_{\sN}'^{2}(n_{\mu})
    +
  \right.
  \\
  &   &
  \hspace{2.0em}
  \left.
    \rule{0em}{3ex}
    +
    \lambda
    v_{0}
    \left[\vec{\varphi}'(n_{\mu})\cdot\vec{\varphi}'(n_{\mu})\right]
    \varphi_{\sN}'(n_{\mu})
    +
    \frac{\lambda}{4}
    \left[\vec{\varphi}'(n_{\mu})\cdot\vec{\varphi}'(n_{\mu})\right]^{2}
  \right\}.
\end{eqnarray*}

\noindent
There are field-odd and field-even terms in this action. Since the
expectation values will single out one of these parities, it is convenient
to write explicitly the field-odd and field-even parts of the non-Gaussian
part of the action,

\noindent
\begin{eqnarray*}
  S_{V,{\rm odd}}[\vec{\varphi}']
  & = &
  \sum_{n_{\mu}}^{N^{d}}
  \left\{
    \rule{0em}{3.0ex}
    v_{0}
    \left[\alpha+\lambda v_{0}^{2}\right]
    \varphi_{\sN}'(n_{\mu})
    -
    j_{0}\varphi_{\sN}'(n_{\mu})
    +
  \right.
  \\
  &   &
  \hspace{2.0em}
  \left.
    \rule{0em}{3.0ex}
    +
    \lambda
    v_{0}
    \left[\vec{\varphi}'(n_{\mu})\cdot\vec{\varphi}'(n_{\mu})\right]
    \varphi_{\sN}'(n_{\mu})
  \right\},
  \\
  S_{V,{\rm even}}[\vec{\varphi}']
  & = &
  \sum_{n_{\mu}}^{N^{d}}
  \left\{
    \frac{\alpha-\alpha_{0}+\lambda v_{0}^{2}}{2}
    \left[\vec{\varphi}'(n_{\mu})\cdot\vec{\varphi}'(n_{\mu})\right]
    +
  \right.
  \\
  &   &
  \hspace{2.0em}
  \left.
    +
    \frac{\alpha_{0}-\alpha_{\sN}+2\lambda v_{0}^{2}}{2}\,
    \varphi_{\sN}'^{2}(n_{\mu})
    +
  \right.
  \\
  &   &
  \hspace{2.0em}
  \left.
    +
    \frac{\lambda}{4}
    \left[\vec{\varphi}'(n_{\mu})\cdot\vec{\varphi}'(n_{\mu})\right]^{2}
  \right\}.
\end{eqnarray*}

\noindent
It is also convenient, for use in the calculations, to write versions of
these expressions in which the terms containing the
$\varphi_{\sN}'(n_{\mu})$ field component are written explicitly, and one
version in which the terms containing the $\varphi_{1}'(n_{\mu})$ field
component are written explicitly as well, rather than as part of the
scalar products,

\noindent
\begin{eqnarray}
  \label{SVodd}
  S_{V,{\rm odd}}[\vec{\varphi}']
  & = &
  \sum_{n_{\mu}}^{N^{d}}
  \left\{
    \rule{0em}{4.0ex}
    v_{0}
    \left[\alpha+\lambda v_{0}^{2}\right]
    \varphi_{\sN}'(n_{\mu})
    -
    j_{0}\varphi_{\sN}'(n_{\mu})
    +
  \right.
  \nonumber\\
  &   &
  \hspace{2.0em}
  \left.
    \rule{0em}{4.0ex}
    +
    \lambda
    v_{0}
    \left[
      \sum_{i=1}^{\sN-1}
      \varphi_{i}'^{2}(n_{\mu})
    \right]
    \varphi_{\sN}'(n_{\mu})
    +
    \lambda
    v_{0}
    \varphi_{\sN}'^{3}(n_{\mu})
  \right\},
  \\
  \label{SVeven1}
  S_{V,{\rm even}}[\vec{\varphi}']
  & = &
  \sum_{n_{\mu}}^{N^{d}}
  \left\{
    \rule{0em}{5ex}
    \frac{\alpha-\alpha_{0}+\lambda v_{0}^{2}}{2}
    \sum_{i=1}^{\sN-1}
    \varphi_{i}'^{2}(n_{\mu})
    +
  \right.
  \nonumber\\
  &   &
  \hspace{2.2em}
  \left.
    +
    \frac{\alpha-\alpha_{\sN}+3\lambda v_{0}^{2}}{2}\,
    \varphi_{\sN}'^{2}(n_{\mu})
    +
  \right.
  \\
  &   &
  \hspace{2.0em}
  \left.
    \rule{0em}{5ex}
    +
    \frac{\lambda}{4}
    \left[
      \sum_{i=1}^{\sN-1}
      \varphi_{i}'^{2}(n_{\mu})
    \right]^{2}
    +
    \frac{\lambda}{2}
    \left[
      \sum_{i=1}^{\sN-1}
      \varphi_{i}'^{2}(n_{\mu})
    \right]
    \varphi_{\sN}'^{2}(n_{\mu})
    +
    \frac{\lambda}{4}\,
    \varphi_{\sN}'^{4}(n_{\mu})
  \right\},
  \nonumber\\
  \label{SVeven2}
  S_{V,{\rm even}}[\vec{\varphi}']
  & = &
  \sum_{n_{\mu}}^{N^{d}}
  \left\{
    \rule{0em}{5ex}
    \frac{\alpha-\alpha_{0}+\lambda v_{0}^{2}}{2}\,
    \varphi_{1}'^{2}(n_{\mu})
    +
  \right.
  \nonumber\\
  &   &
  \hspace{2.2em}
  \left.
    +
    \frac{\alpha-\alpha_{0}+\lambda v_{0}^{2}}{2}
    \sum_{i=2}^{\sN-1}
    \varphi_{i}'^{2}(n_{\mu})
    +
  \right.
  \nonumber\\
  &   &
  \hspace{2.2em}
  \left.
    +
    \frac{\alpha-\alpha_{\sN}+3\lambda v_{0}^{2}}{2}\,
    \varphi_{\sN}'^{2}(n_{\mu})
    +
  \right.
  \\
  &   &
  \hspace{2.0em}
  \left.
    \rule{0em}{5ex}
    +
    \frac{\lambda}{4}\,
    \varphi_{1}'^{4}(n_{\mu})
    +
    \frac{\lambda}{2}\,
    \varphi_{1}'^{2}(n_{\mu})
    \left[
      \sum_{i=2}^{\sN-1}
      \varphi_{i}'^{2}(n_{\mu})
    \right]
    +
    \frac{\lambda}{4}
    \left[
      \sum_{i=2}^{\sN-1}
      \varphi_{i}'^{2}(n_{\mu})
    \right]^{2}
    +
  \right.
  \nonumber\\
  &   &
  \hspace{2.0em}
  \left.
    \rule{0em}{5ex}
    +
    \frac{\lambda}{2}\,
    \varphi_{1}'^{2}(n_{\mu})
    \varphi_{\sN}'^{2}(n_{\mu})
    +
    \frac{\lambda}{2}
    \left[
      \sum_{i=2}^{\sN-1}
      \varphi_{i}'^{2}(n_{\mu})
    \right]
    \varphi_{\sN}'^{2}(n_{\mu})
    +
    \frac{\lambda}{4}\,
    \varphi_{\sN}'^{4}(n_{\mu})
  \right\}.
  \nonumber
\end{eqnarray}

\subsection{The Expectation Value of \boldmath
  $\varphi_{\sN}'(n_{\mu}')S_{V}[\vec{\varphi}']$}

Let us now calculate the expectation value

\begin{displaymath}
  \left\langle
    \varphi_{\sN}'(n_{\mu}')S_{V}[\vec{\varphi}']
  \right\rangle_{0}.
\end{displaymath}

\noindent
Since $S_{0}[\vec{\varphi}']$ is field-even, all expectation values of
field-odd observables are zero when calculated in its measure. Therefore
it is necessary that the observables be field-even if their expectation
values are to be non-zero. Since in this case we have an explicit factor
of $\varphi_{\sN}'(n_{\mu}')$, it follows that only the field-odd part of
$S_{V}[\vec{\varphi}']$ will contribute to this expectation value,

\begin{displaymath}
  \left\langle
    \varphi_{\sN}'(n_{\mu}')S_{V}[\vec{\varphi}']
  \right\rangle_{0}
  =
  \left\langle
    \varphi_{\sN}'(n_{\mu}')S_{V,{\rm odd}}[\vec{\varphi}']
  \right\rangle_{0}.
\end{displaymath}

\noindent
If we write the expectation value out, using the form of the action
$S_{V,{\rm odd}}[\vec{\varphi}']$ given in Equation~(\ref{SVodd}), we get

\noindent
\begin{eqnarray*}
  \left\langle
    \varphi_{\sN}'(n_{\mu}')S_{V}[\vec{\varphi}']
  \right\rangle_{0}
  & = &
  \sum_{n_{\mu}}^{N^{d}}
  \left\{
    \rule{0em}{4ex}
    v_{0}
    \left[\alpha+\lambda v_{0}^{2}\right]
    \left\langle
      \varphi_{\sN}'(n_{\mu})\varphi_{\sN}'(n_{\mu}')
    \right\rangle_{0}
    +
  \right.
  \\
  &   &
  \hspace{2.2em}
  \left.
    -
    j_{0}
    \left\langle
      \varphi_{\sN}'(n_{\mu})\varphi_{\sN}'(n_{\mu}')    
    \right\rangle_{0}
    +
  \right.
  \\
  &   &
  \hspace{2.0em}
  \left.
    \rule{0em}{4ex}
    +
    \lambda
    v_{0}
    \sum_{i=1}^{\sN-1}
    \left\langle
      \varphi_{i}'^{2}(n_{\mu})
    \right\rangle_{0}
    \left\langle
      \varphi_{\sN}'(n_{\mu})\varphi_{\sN}'(n_{\mu}')
    \right\rangle_{0}
    +
  \right.
  \\
  &   &
  \hspace{2.0em}
  \left.
    \rule{0em}{4ex}
    +
    \lambda
    v_{0}
    \left\langle
      \varphi_{\sN}'^{3}(n_{\mu})\varphi_{\sN}'(n_{\mu}')
    \right\rangle_{0}
  \right\}.
\end{eqnarray*}

\noindent
The expectation values in the first three terms turn out to be just the
position-space propagator for the $\varphi_{\sN}'(n_{\mu})$ field
component. From Appendix~\ref{TabofInts}, Equation~(\ref{SinglePropNX}),
we get

\begin{displaymath}
  g_{\sN}(n_{\mu}-n_{\mu}')
  =
  \frac{1}{N^{d}}
  \sum_{k_{\mu}}^{N^{d}}
  \frac
  {e^{-\ii(2\pi/N)\sum_{\mu}^{d}k_{\mu}(n_{\mu}-n_{\mu}')}}
  {\rho^{2}(k_{\mu})+\alpha_{\sN}},
\end{displaymath}

\noindent
which is just the statement that $g_{\sN}(n_{\mu}-n_{\mu}')$ is the
inverse Fourier transform of the momentum-space propagator. We also have
the corresponding result for the other field $\sN-1$ components, with
$i\neq\sN$,

\noindent
\begin{eqnarray*}
  g_{0}(n_{\mu}-n_{\mu}')
  & = &
  \left\langle
    \varphi_{i}'(n_{\mu})\varphi_{i}'(n_{\mu}')
  \right\rangle_{0}
  \\
  & = &
  \frac{1}{N^{d}}
  \sum_{k_{\mu}}^{N^{d}}
  \frac
  {e^{-\ii(2\pi/N)\sum_{\mu}^{d}k_{\mu}(n_{\mu}-n_{\mu}')}}
  {\rho^{2}(k_{\mu})+\alpha_{0}}.
\end{eqnarray*}

\noindent
The second expectation value that we must calculate, with $i\neq\sN$, is
simply

\noindent
\begin{eqnarray*}
  g_{0}(0)
  & = &
  \left\langle
    \varphi_{i}'^{2}(n_{\mu})
  \right\rangle_{0}
  \\
  & = &
  \sigma_{0}^{2}
  \\
  & = &
  \frac{1}{N^{d}}
  \sum_{k_{\mu}}^{N^{d}}
  \frac{1}{\rho^{2}(k_{\mu})+\alpha_{0}}.
\end{eqnarray*}

\noindent
The third expectation value that we must calculate can be found in
Appendix~\ref{TabofInts}, Equation~(\ref{DecompFN3FN1}), and can be shown
to be given in terms of the first one by

\begin{displaymath}
  \left\langle
    \varphi_{\sN}'^{3}(n_{\mu})\varphi_{\sN}'(n_{\mu}')
  \right\rangle_{0}
  =
  3\sigma_{\sN}^{2}\,g_{\sN}(n_{\mu}-n_{\mu}').
\end{displaymath}

\noindent
We are thus left with a simpler form for the expectation value,

\noindent
\begin{eqnarray*}
  \left\langle
    \varphi_{\sN}'(n_{\mu}')S_{V}[\vec{\varphi}']
  \right\rangle_{0}
  & = &
  \sum_{n_{\mu}}^{N^{d}}
  \left\{
    \rule{0em}{4ex}
    v_{0}
    \alpha\,
    g_{\sN}(n_{\mu}-n_{\mu}')
    +
  \right.
  \\
  &   &
  \hspace{2.2em}
  \left.
    +
    v_{0}^{3}
    \lambda\,
    g_{\sN}(n_{\mu}-n_{\mu}')
    +
  \right.
  \\
  &   &
  \hspace{2.0em}
  \left.
    \rule{0em}{4ex}
    +
    v_{0}
    (\sN-1)
    \lambda
    \sigma_{0}^{2}\,
    g_{\sN}(n_{\mu}-n_{\mu}')
    +
  \right.
  \\
  &   &
  \hspace{2.0em}
  \left.
    \rule{0em}{4ex}
    +
    v_{0}
    3\lambda
    \sigma_{\sN}^{2}\,
    g_{\sN}(n_{\mu}-n_{\mu}')
    +
  \right.
  \\
  &   &
  \hspace{2.0em}
  \left.
    \rule{0em}{4ex}
    -
    j_{0}
    g_{\sN}(n_{\mu}-n_{\mu}')
  \right\}.
\end{eqnarray*}

\noindent
In all terms the only quantity still depending on $n_{\mu}$ is
$g_{\sN}(n_{\mu}-n_{\mu}')$, so that we can write this equation as

\begin{displaymath}
  \left\langle
    \varphi_{\sN}'(n_{\mu}')S_{V}[\vec{\varphi}']
  \right\rangle_{0}
  =
  \left\{
    v_{0}
    \left[
      \alpha
      +
      v_{0}^{2}
      \lambda
      +
      (\sN-1)
      \lambda
      \sigma_{0}^{2}
      +
      3\lambda
      \sigma_{\sN}^{2}
    \right]
    -
    j_{0}
  \right\}
  \sum_{n_{\mu}}^{N^{d}}
  g_{\sN}(n_{\mu}-n_{\mu}').
\end{displaymath}

\noindent
Using Equation~(\ref{SumgN}), which gives this final sum, we may finally
write

\begin{equation}\label{ExpecValFNSV}
  \left\langle
    \varphi_{\sN}'(n_{\mu}')S_{V}[\vec{\varphi}']
  \right\rangle_{0}
  =
  \frac
  {
    v_{0}
    \left[
      \alpha
      +
      v_{0}^{2}
      \lambda
      +
      (\sN-1)
      \lambda
      \sigma_{0}^{2}
      +
      3\lambda
      \sigma_{\sN}^{2}
    \right]
    -
    j_{0}
  }
  {\alpha_{\sN}}.
\end{equation}

\subsection{The Expectation Value of \boldmath $S_{V}[\vec{\varphi}']$}

We now calculate the expectation value

\begin{displaymath}
  \left\langle
    S_{V}[\vec{\varphi}']
  \right\rangle_{0}.
\end{displaymath}

\noindent
Only the field-even part of the action will yield a non-zero result, so
that we have

\begin{displaymath}
  \left\langle
    S_{V}[\vec{\varphi}']
  \right\rangle_{0}
  =
  \left\langle
    S_{V,{\rm even}}[\vec{\varphi}']
  \right\rangle_{0}.
\end{displaymath}

\noindent
Using the form of $S_{V,{\rm even}}[\vec{\varphi}']$ shown in
Equation~(\ref{SVeven1}) we get for this expectation value

\noindent
\begin{eqnarray*}
  \lefteqn
  {
    \left\langle
      S_{V}[\vec{\varphi}']
    \right\rangle_{0}
  }
  &   &
  \\
  & = &
  \sum_{n_{\mu}}^{N^{d}}
  \left\{
    \rule{0em}{5ex}
    \frac{\alpha-\alpha_{0}+\lambda v_{0}^{2}}{2}
    \sum_{i=1}^{\sN-1}
    \left\langle
      \varphi_{i}'^{2}(n_{\mu})
    \right\rangle_{0}
    +
  \right.
  \\
  &   &
  \hspace{2.45em}
  \left.
    +
    \frac{\alpha-\alpha_{\sN}+3\lambda v_{0}^{2}}{2}\,
    \left\langle
      \varphi_{\sN}'^{2}(n_{\mu})
    \right\rangle_{0}
    +
  \right.
  \\
  &   &
  \hspace{2.2em}
  \left.
    \rule{0em}{5ex}
    +
    \frac{\lambda}{4}
    \left\langle
      \left[
        \sum_{i=1}^{\sN-1}
        \varphi_{i}'^{2}(n_{\mu})
      \right]^{2}
    \right\rangle_{0}
    +
    \frac{\lambda}{2}
    \left[
      \sum_{i=1}^{\sN-1}
      \left\langle
        \varphi_{i}'^{2}(n_{\mu})
      \right\rangle_{0}
    \right]
    \left\langle
      \varphi_{\sN}'^{2}(n_{\mu})
    \right\rangle_{0}
    +
    \frac{\lambda}{4}\,
    \left\langle
      \varphi_{\sN}'^{4}(n_{\mu})
    \right\rangle_{0}
  \right\}.
\end{eqnarray*}

\noindent
Most of the remaining expectation values can be written in terms of
$\sigma_{0}$ and $\sigma_{\sN}$, if we recall that it can be shown that
for $i\neq\sN$ we have

\begin{displaymath}
  \left\langle
    \varphi_{i}'^{4}(n_{\mu})
  \right\rangle_{0}
  =
  3\sigma_{0}^{2},
\end{displaymath}

\noindent
while for $i=\sN$ we have, in a similar way,

\begin{displaymath}
  \left\langle
    \varphi_{\sN}'^{4}(n_{\mu})
  \right\rangle_{0}
  =
  3\sigma_{\sN}^{2},
\end{displaymath}

\noindent
as one can find in Appendix~\ref{TabofInts}, Equation~(\ref{DecompFN4}).
Given all this, we may write for our expectation value

\noindent
\begin{eqnarray*}
  \left\langle
    S_{V}[\vec{\varphi}']
  \right\rangle_{0}
  & = &
  \sum_{n_{\mu}}^{N^{d}}
  \left\{
    \rule{0em}{5ex}
    \frac{\alpha-\alpha_{0}+\lambda v_{0}^{2}}{2}\,
    (\sN-1)
    \sigma_{0}^{2}
    +
    \frac{\alpha-\alpha_{\sN}+3\lambda v_{0}^{2}}{2}\,
    \sigma_{\sN}^{2}
    +
  \right.
  \\
  &   &
  \hspace{2.2em}
  \left.
    +
    \frac{\lambda}{4}
    \left\langle
      \left[
        \sum_{i=1}^{\sN-1}
        \varphi_{i}'^{2}(n_{\mu})
      \right]^{2}
    \right\rangle_{0}
    +
    \frac{\lambda}{2}\,
    (\sN-1)
    \sigma_{0}^{2}
    \sigma_{\sN}^{2}
    +
    \frac{3\lambda}{4}\,
    \sigma_{\sN}^{4}
  \right\}.
\end{eqnarray*}

\noindent
The remaining expectation value of the sum shown can be found in
Appendix~\ref{TabofInts}, Equation~(\ref{SumFiN-1}),

\begin{displaymath}
  \left\langle
    \left[
      \sum_{i=1}^{\sN-1}
      \varphi_{i}'^{2}(n_{\mu})
    \right]^{2}
  \right\rangle_{0}
  =
  \rule{0em}{4ex}
  (\sN+1)(\sN-1)
  \sigma_{0}^{4}.
\end{displaymath}

\noindent
Using this result we get for our expectation value

\noindent
\begin{eqnarray*}
  \left\langle
    S_{V}[\vec{\varphi}']
  \right\rangle_{0}
  & = &
  \sum_{n_{\mu}}^{N^{d}}
  \left[
    \rule{0em}{5ex}
    \frac{\alpha-\alpha_{0}+\lambda v_{0}^{2}}{2}\,
    (\sN-1)
    \sigma_{0}^{2}
    +
    \frac{\alpha-\alpha_{\sN}+3\lambda v_{0}^{2}}{2}\,
    \sigma_{\sN}^{2}
    +
  \right.
  \\
  &   &
  \hspace{2.2em}
  \left.
    \rule{0em}{5ex}
    +
    \frac{\lambda}{4}\,
    (\sN^{2}-1)
    \sigma_{0}^{4}
    +
    \frac{\lambda}{2}\,
    (\sN-1)
    \sigma_{0}^{2}
    \sigma_{\sN}^{2}
    +
    \frac{3\lambda}{4}\,
    \sigma_{\sN}^{4}
  \right].
\end{eqnarray*}

\noindent
Note that all the sums can now be done, so that we can write our result in
the simpler form

\noindent
\begin{eqnarray}\label{ExpecValSV}
  \left\langle
    S_{V}[\vec{\varphi}']
  \right\rangle_{0}
  & = &
  N^{d}
  \left[
    \frac{\alpha-\alpha_{0}+\lambda v_{0}^{2}}{2}\,
    (\sN-1)
    \sigma_{0}^{2}
    +
    \frac{\alpha-\alpha_{\sN}+3\lambda v_{0}^{2}}{2}\,
    \sigma_{\sN}^{2}
    +
  \right.
  \nonumber\\
  &   &
  \hspace{2.2em}
  \left.
    +
    \frac{\lambda}{4}\,
    (\sN^{2}-1)
    \sigma_{0}^{4}
    +
    \frac{\lambda}{2}\,
    (\sN-1)
    \sigma_{0}^{2}
    \sigma_{\sN}^{2}
    +
    \frac{3\lambda}{4}\,
    \sigma_{\sN}^{4}
  \right].
\end{eqnarray}

\subsection{The Expectation Value of \boldmath
  $\varphi_{1}'(n_{\mu}')\varphi_{1}'(n_{\mu}'')S_{V}[\vec{\varphi}']$}

We now calculate the expectation value

\begin{displaymath}
  \left\langle
    \varphi_{1}'(n_{\mu}')
    \varphi_{1}'(n_{\mu}'')
    S_{V}[\vec{\varphi}']
  \right\rangle_{0}.
\end{displaymath}

\noindent
Once more only the field-even part of the action will yield a non-zero
result, so that we have

\begin{displaymath}
  \left\langle
    \varphi_{1}'(n_{\mu}')
    \varphi_{1}'(n_{\mu}'')
    S_{V}[\vec{\varphi}']
  \right\rangle_{0}
  =
  \left\langle
    \varphi_{1}'(n_{\mu}')
    \varphi_{1}'(n_{\mu}'')
    S_{V,{\rm even}}[\vec{\varphi}']
  \right\rangle_{0}.
\end{displaymath}

\noindent
Using the form of $S_{V,{\rm even}}[\vec{\varphi}']$ shown in
Equation~(\ref{SVeven2}), and if we already replace the expectation values
of squared fields by $\sigma_{0}$ or $\sigma_{\sN}$ whenever possible, as
well as replace
$\left\langle\varphi_{1}'(n_{\mu}')\varphi_{1}'(n_{\mu}'')\right\rangle_{0}$
by $g_{0}(n_{\mu}'-n_{\mu}'')$, we get

\noindent
\begin{eqnarray*}
  \left\langle
    \varphi_{1}'(n_{\mu}')
    \varphi_{1}'(n_{\mu}'')
    S_{V}[\vec{\varphi}']
  \right\rangle_{0}
  & = &
  \sum_{n_{\mu}}^{N^{d}}
  \left\{
    \rule{0em}{5ex}
    \frac{\alpha-\alpha_{0}+\lambda v_{0}^{2}}{2}\,
    \left\langle
      \varphi_{1}'^{2}(n_{\mu})
      \varphi_{1}'(n_{\mu}')
      \varphi_{1}'(n_{\mu}'')
    \right\rangle_{0}
    +
  \right.
  \\
  &   &
  \hspace{2.45em}
  \left.
    +
    \frac{\alpha-\alpha_{0}+\lambda v_{0}^{2}}{2}\,
    (\sN-2)
    \sigma_{0}^{2}\,
    g_{0}(n_{\mu}'-n_{\mu}'')
    +
  \right.
  \\
  &   &
  \hspace{2.2em}
  \left.
    \rule{0em}{5ex}
    +
    \frac{\alpha-\alpha_{\sN}+3\lambda v_{0}^{2}}{2}\,
    \sigma_{\sN}^{2}\,
    g_{0}(n_{\mu}'-n_{\mu}'')
    +
  \right.
  \\
  &   &
  \hspace{2.2em}
  \left.
    \rule{0em}{5ex}
    +
    \frac{\lambda}{4}\,
    \left\langle
      \varphi_{1}'^{4}(n_{\mu})
      \varphi_{1}'(n_{\mu}')
      \varphi_{1}'(n_{\mu}'')
    \right\rangle_{0}
    +
  \right.
  \\
  &   &
  \hspace{2.2em}
  \left.
    \rule{0em}{5ex}
    +
    \frac{\lambda}{2}\,
    (\sN-2)
    \sigma_{0}^{2}
    \left\langle
      \varphi_{1}'^{2}(n_{\mu})
      \varphi_{1}'(n_{\mu}')
      \varphi_{1}'(n_{\mu}'')
    \right\rangle_{0}
    +
  \right.
  \\
  &   &
  \hspace{2.2em}
  \left.
    \rule{0em}{5ex}
    +
    \frac{\lambda}{2}\,
    \sigma_{\sN}^{2}
    \left\langle
      \varphi_{1}'^{2}(n_{\mu})
      \varphi_{1}'(n_{\mu}')
      \varphi_{1}'(n_{\mu}'')
    \right\rangle_{0}
    +
  \right.
  \\
  &   &
  \hspace{2.2em}
  \left.
    \rule{0em}{5ex}
    +
    \frac{\lambda}{2}
    (\sN-2)
    \sigma_{0}^{2}
    \sigma_{\sN}^{2}\,
    g_{0}(n_{\mu}'-n_{\mu}'')
    +
  \right.
  \\
  &   &
  \hspace{2.2em}
  \left.
    \rule{0em}{5ex}
    +
    \frac{\lambda}{4}
    \left\langle
      \left[
        \sum_{i=2}^{\sN-1}
        \varphi_{i}'^{2}(n_{\mu})
      \right]^{2}
    \right\rangle_{0}
    g_{0}(n_{\mu}'-n_{\mu}'')
    +
  \right.
  \\
  &   &
  \hspace{2.2em}
  \left.
    \rule{0em}{5ex}
    +
    \frac{\lambda}{4}\,
    \left\langle
      \varphi_{\sN}'^{4}(n_{\mu})
    \right\rangle_{0}
    g_{0}(n_{\mu}'-n_{\mu}'')
  \right\}.
\end{eqnarray*}

\noindent
We may now use the known value of the expectation value of the squared
sum. From Appendix~\ref{TabofInts}, Equation~(\ref{SumFiN-2}), we get

\begin{displaymath}
  \left\langle
    \left[
      \sum_{i=2}^{\sN-1}
      \varphi_{i}'^{2}(n_{\mu})
    \right]^{2}
  \right\rangle_{0}
  =
  \sN(\sN-2)
  \sigma_{0}^{4}.
\end{displaymath}

\noindent
We may also use the fact that it can be shown that

\noindent
\begin{eqnarray*}
  \left\langle
    \varphi_{\sN}'^{4}(n_{\mu})
  \right\rangle_{0}
  & = &
  3\sigma_{\sN}^{4},
  \\
  \left\langle
    \varphi_{1}'^{2}(n_{\mu})
    \varphi_{1}'(n_{\mu}')
    \varphi_{1}'(n_{\mu}'')
  \right\rangle_{0}
  & = &
  \sigma_{0}^{2}\,
  g_{0}(n_{\mu}'-n_{\mu}'')
  +
  2\,
  g_{0}(n_{\mu}-n_{\mu}')\,
  g_{0}(n_{\mu}-n_{\mu}''),
  \\
  \left\langle
    \varphi_{1}'^{4}(n_{\mu})
    \varphi_{1}'(n_{\mu}')
    \varphi_{1}'(n_{\mu}'')
  \right\rangle_{0}
  & = &
  3\sigma_{0}^{4}\,
  g_{0}(n_{\mu}'-n_{\mu}'')
  +
  12\sigma_{0}^{2}\,
  g_{0}(n_{\mu}-n_{\mu}')\,
  g_{0}(n_{\mu}-n_{\mu}''),
\end{eqnarray*}

\noindent
also found in Appendix~\ref{TabofInts}, Equations~(\ref{DecompFN4}),
(\ref{DecompFi2Fi1Fi1}) and~(\ref{DecompFi4Fi1Fi1}), in order to write for
our expectation value

\noindent
\begin{eqnarray*}
  \left\langle
    \varphi_{1}'(n_{\mu}')
    \varphi_{1}'(n_{\mu}'')
    S_{V}[\vec{\varphi}']
  \right\rangle_{0}
  & = &
  \sum_{n_{\mu}}^{N^{d}}
  \left\{
    \rule{0em}{5ex}
    \frac{\alpha-\alpha_{0}+\lambda v_{0}^{2}}{2}\,
    \sigma_{0}^{2}\,
    g_{0}(n_{\mu}'-n_{\mu}'')
    +
  \right.
  \\
  &   &
  \hspace{2.4em}
  \left.
    +
    \left[\alpha-\alpha_{0}+\lambda v_{0}^{2}\right]
    g_{0}(n_{\mu}-n_{\mu}')\,
    g_{0}(n_{\mu}-n_{\mu}'')
    +
  \right.
  \\
  &   &
  \hspace{2.2em}
  \left.
    \rule{0em}{5ex}
    +
    \frac{\alpha-\alpha_{0}+\lambda v_{0}^{2}}{2}\,
    (\sN-2)
    \sigma_{0}^{2}\,
    g_{0}(n_{\mu}'-n_{\mu}'')
    +
  \right.
  \\
  &   &
  \hspace{2.2em}
  \left.
    \rule{0em}{5ex}
    +
    \frac{\alpha-\alpha_{\sN}+3\lambda v_{0}^{2}}{2}\,
    \sigma_{\sN}^{2}\,
    g_{0}(n_{\mu}'-n_{\mu}'')
    +
  \right.
  \\
  &   &
  \hspace{2.2em}
  \left.
    \rule{0em}{5ex}
    +
    \frac{3\lambda}{4}\,
    \sigma_{0}^{4}\,
    g_{0}(n_{\mu}'-n_{\mu}'')
    +
  \right.
  \\
  &   &
  \hspace{2.2em}
  \left.
    \rule{0em}{5ex}
    +
    3\lambda
    \sigma_{0}^{2}\,
    g_{0}(n_{\mu}-n_{\mu}')\,
    g_{0}(n_{\mu}-n_{\mu}'')
    +
  \right.
  \\
  &   &
  \hspace{2.2em}
  \left.
    \rule{0em}{5ex}
    +
    \frac{\lambda}{2}\,
    (\sN-2)
    \sigma_{0}^{4}\,
    g_{0}(n_{\mu}'-n_{\mu}'')
    +
  \right.
  \\
  &   &
  \hspace{2.2em}
  \left.
    \rule{0em}{5ex}
    +
    \lambda
    (\sN-2)
    \sigma_{0}^{2}\,
    g_{0}(n_{\mu}-n_{\mu}')\,
    g_{0}(n_{\mu}-n_{\mu}'')
    +
  \right.
  \\
  &   &
  \hspace{2.2em}
  \left.
    \rule{0em}{5ex}
    +
    \frac{\lambda}{2}\,
    \sigma_{\sN}^{2}
    \sigma_{0}^{2}\,
    g_{0}(n_{\mu}'-n_{\mu}'')
    +
  \right.
  \\
  &   &
  \hspace{2.2em}
  \left.
    \rule{0em}{5ex}
    +
    \lambda
    \sigma_{\sN}^{2}\,
    g_{0}(n_{\mu}-n_{\mu}')\,
    g_{0}(n_{\mu}-n_{\mu}'')
    +
  \right.
  \\
  &   &
  \hspace{2.2em}
  \left.
    \rule{0em}{5ex}
    +
    \frac{\lambda}{2}
    (\sN-2)
    \sigma_{0}^{2}
    \sigma_{\sN}^{2}\,
    g_{0}(n_{\mu}'-n_{\mu}'')
    +
  \right.
  \\
  &   &
  \hspace{2.2em}
  \left.
    \rule{0em}{5ex}
    +
    \frac{\lambda}{4}\,
    \sN(\sN-2)
    \sigma_{0}^{4}\,
    g_{0}(n_{\mu}'-n_{\mu}'')
    +
  \right.
  \\
  &   &
  \hspace{2.2em}
  \left.
    \rule{0em}{5ex}
    +
    \frac{3\lambda}{4}\,
    \sigma_{\sN}^{4}\,
    g_{0}(n_{\mu}'-n_{\mu}'')
  \right\}.
\end{eqnarray*}

\noindent
Next we group all terms containing $g_{0}(n_{\mu}'-n_{\mu}'')$ and
simplify to get

\noindent
\begin{eqnarray*}
  \lefteqn
  {
    \left\langle
      \varphi_{1}'(n_{\mu}')
      \varphi_{1}'(n_{\mu}'')
      S_{V}[\vec{\varphi}']
    \right\rangle_{0}
  }
  &   &
  \\
  & = &
  \sum_{n_{\mu}}^{N^{d}}
  \left[
    \rule{0em}{4ex}
    \frac{\alpha-\alpha_{0}+\lambda v_{0}^{2}}{2}\,
    (\sN-1)
    \sigma_{0}^{2}
    +
    \frac{\alpha-\alpha_{\sN}+3\lambda v_{0}^{2}}{2}\,
    \sigma_{\sN}^{2}
    +
  \right.
  \\
  &   &
  \hspace{2.2em}
  \left.
    \rule{0em}{4ex}
    +
    \frac{\lambda}{4}\,
    (\sN^{2}-1)
    \sigma_{0}^{4}
    +
    \frac{\lambda}{2}\,
    (\sN-1)
    \sigma_{0}^{2}
    \sigma_{\sN}^{2}
    +
    \frac{3\lambda}{4}\,
    \sigma_{\sN}^{4}
  \right]
  g_{0}(n_{\mu}'-n_{\mu}'')
  \\
  &   &
  +
  \sum_{n_{\mu}}^{N^{d}}
  \left\{
    \rule{0em}{3ex}
    \left[
      \alpha-\alpha_{0}+\lambda v_{0}^{2}
    \right]
    +
    \lambda
    (\sN+1)
    \sigma_{0}^{2}
    +
    \lambda
    \sigma_{\sN}^{2}
  \right\}
  g_{0}(n_{\mu}-n_{\mu}')\,
  g_{0}(n_{\mu}-n_{\mu}'').
\end{eqnarray*}

\noindent
The sum over $n_{\mu}$ can now be done in all terms in the first group,
yielding

\noindent
\begin{eqnarray*}
  \lefteqn
  {
    \left\langle
      \varphi_{1}'(n_{\mu}')
      \varphi_{1}'(n_{\mu}'')
      S_{V}[\vec{\varphi}']
    \right\rangle_{0}
  }
  &   &
  \\
  & = &
  N^{d}
  \left[
    \frac{\alpha-\alpha_{0}+\lambda v_{0}^{2}}{2}\,
    (\sN-1)
    \sigma_{0}^{2}
    +
    \frac{\alpha-\alpha_{\sN}+3\lambda v_{0}^{2}}{2}\,
    \sigma_{\sN}^{2}
    +
  \right.
  \\
  &   &
  \hspace{2.2em}
  \left.
    +
    \frac{\lambda}{4}\,
    (\sN^{2}-1)
    \sigma_{0}^{4}
    +
    \frac{\lambda}{2}\,
    (\sN-1)
    \sigma_{0}^{2}
    \sigma_{\sN}^{2}
    +
    \frac{3\lambda}{4}\,
    \sigma_{\sN}^{4}
  \right]
  g_{0}(n_{\mu}'-n_{\mu}'')
  \\
  &   &
  +
  \left[
    \alpha-\alpha_{0}
    +
    \lambda
    v_{0}^{2}
    +
    \lambda
    (\sN+1)
    \sigma_{0}^{2}
    +
    \lambda
    \sigma_{\sN}^{2}
  \right]
  \sum_{n_{\mu}}^{N^{d}}
  g_{0}(n_{\mu}-n_{\mu}')\,
  g_{0}(n_{\mu}-n_{\mu}'').
\end{eqnarray*}

\noindent
We must now perform the sum indicated. This is easily done using Fourier
transforms. From Appendix~\ref{TabofInts}, Equation~(\ref{DualProp0X}), we
get

\begin{displaymath}
  \sum_{n_{\mu}}^{N^{d}}
  g_{0}(n_{\mu}-n_{\mu}')\,
  g_{0}(n_{\mu}-n_{\mu}'')
  =
  \frac{1}{N^{d}}
  \sum_{k_{\mu}}^{N^{d}}
  \frac
  {
    e^{-\ii(2\pi/N)\sum_{\mu}^{d}k_{\mu}(n_{\mu}'-n_{\mu}'')}
  }
  {
    [\rho^{2}(k_{\mu})+\alpha_{0}]^{2}
  },
\end{displaymath}

\noindent
which is expressed as a Fourier transform, with the general structure of a
two-point function. We have therefore the final result,

\noindent
\begin{eqnarray}\label{ExpecValF1F1SV}
  \lefteqn
  {
    \left\langle
      \varphi_{1}'(n_{\mu}')
      \varphi_{1}'(n_{\mu}'')
      S_{V}[\vec{\varphi}']
    \right\rangle_{0}
  }
  &   &
  \nonumber\\
  & = &
  N^{d}
  \left[
    \frac{\alpha-\alpha_{0}+\lambda v_{0}^{2}}{2}\,
    (\sN-1)
    \sigma_{0}^{2}
    +
    \frac{\alpha-\alpha_{\sN}+3\lambda v_{0}^{2}}{2}\,
    \sigma_{\sN}^{2}
    +
  \right.
  \nonumber\\
  &   &
  \hspace{2.2em}
  \left.
    +
    \frac{\lambda}{4}\,
    (\sN^{2}-1)
    \sigma_{0}^{4}
    +
    \frac{\lambda}{2}\,
    (\sN-1)
    \sigma_{0}^{2}
    \sigma_{\sN}^{2}
    +
    \frac{3\lambda}{4}\,
    \sigma_{\sN}^{4}
  \right]
  g_{0}(n_{\mu}'-n_{\mu}'')
  \\
  &   &
  +
  \left[
    \alpha-\alpha_{0}
    +
    \lambda
    v_{0}^{2}
    +
    \lambda
    (\sN+1)
    \sigma_{0}^{2}
    +
    \lambda
    \sigma_{\sN}^{2}
  \right]
  \frac{1}{N^{d}}
  \sum_{k_{\mu}}^{N^{d}}
  \frac
  {
    e^{-\ii(2\pi/N)\sum_{\mu}^{d}k_{\mu}(n_{\mu}'-n_{\mu}'')}
  }
  {
    [\rho^{2}(k_{\mu})+\alpha_{0}]^{2}
  }.
  \nonumber
\end{eqnarray}

\subsection{The Expectation Value of \boldmath
  $\varphi_{\sN}'(n_{\mu}')\varphi_{\sN}'(n_{\mu}'')S_{V}[\vec{\varphi}']$}

We now calculate the expectation value

\begin{displaymath}
  \left\langle
    \varphi_{\sN}'(n_{\mu}')
    \varphi_{\sN}'(n_{\mu}'')
    S_{V}[\vec{\varphi}']
  \right\rangle_{0}.
\end{displaymath}

\noindent
Once again only the field-even part of the action will yield a non-zero
result, so that we have

\begin{displaymath}
  \left\langle
    \varphi_{\sN}'(n_{\mu}')
    \varphi_{\sN}'(n_{\mu}'')
    S_{V}[\vec{\varphi}']
  \right\rangle_{0}
  =
  \left\langle
    \varphi_{\sN}'(n_{\mu}')
    \varphi_{\sN}'(n_{\mu}'')
    S_{V,{\rm even}}[\vec{\varphi}']
  \right\rangle_{0}.
\end{displaymath}

\noindent
Using the form of $S_{V,{\rm even}}[\vec{\varphi}']$ shown in
Equation~(\ref{SVeven1}), and if we already replace the expectation values
of squared fields by $\sigma_{0}$ or $\sigma_{\sN}$ whenever possible, as
well as replace
$\left\langle\varphi_{\sN}'(n_{\mu}')\varphi_{\sN}'(n_{\mu}'')\right\rangle_{0}$
by $g_{\sN}(n_{\mu}'-n_{\mu}'')$, we get

\noindent
\begin{eqnarray*}
  \left\langle
    \varphi_{\sN}'(n_{\mu}')
    \varphi_{\sN}'(n_{\mu}'')
    S_{V}[\vec{\varphi}']
  \right\rangle_{0}
  & = &
  \sum_{n_{\mu}}^{N^{d}}
  \left\{
    \rule{0em}{5ex}
    \frac{\alpha-\alpha_{0}+\lambda v_{0}^{2}}{2}\,
    (\sN-1)
    \sigma_{0}^{2}\,
    g_{\sN}(n_{\mu}'-n_{\mu}'')
    +
  \right.
  \\
  &   &
  \hspace{2.42em}
  \left.
    +
    \frac{\alpha-\alpha_{\sN}+3\lambda v_{0}^{2}}{2}\,
    \left\langle
      \varphi_{\sN}'^{2}(n_{\mu})
      \varphi_{\sN}'(n_{\mu}')
      \varphi_{\sN}'(n_{\mu}'')
    \right\rangle_{0}
    +
  \right.
  \\
  &   &
  \hspace{2.2em}
  \left.
    \rule{0em}{5ex}
    +
    \frac{\lambda}{4}
    \left\langle
      \left[
        \sum_{i=1}^{\sN-1}
        \varphi_{i}'^{2}(n_{\mu})
      \right]^{2}
    \right\rangle_{0}
    g_{\sN}(n_{\mu}'-n_{\mu}'')
    +
  \right.
  \\
  &   &
  \hspace{2.2em}
  \left.
    \rule{0em}{5ex}
    +
    \frac{\lambda}{2}\,
    (\sN-1)
    \sigma_{0}^{2}
    \left\langle
      \varphi_{\sN}'^{2}(n_{\mu})
      \varphi_{\sN}'(n_{\mu}')
      \varphi_{\sN}'(n_{\mu}'')
    \right\rangle_{0}
    +
  \right.
  \\
  &   &
  \hspace{2.2em}
  \left.
    \rule{0em}{5ex}
    +
    \frac{\lambda}{4}\,
    \left\langle
      \varphi_{\sN}'^{4}(n_{\mu})
      \varphi_{\sN}'(n_{\mu}')
      \varphi_{\sN}'(n_{\mu}'')
    \right\rangle_{0}
  \right\}.
\end{eqnarray*}

\noindent
We may now use the known value of the expectation value of the squared
sum, found in Appendix~\ref{TabofInts}, Equation~(\ref{SumFiN-1}),

\begin{displaymath}
  \left\langle
    \left[
      \sum_{i=1}^{\sN-1}
      \varphi_{i}'^{2}(n_{\mu})
    \right]^{2}
  \right\rangle_{0}
  =
  (\sN+1)(\sN-1)
  \sigma_{0}^{4},
\end{displaymath}

\noindent
as well as the fact that it can be shown that

\noindent
\begin{eqnarray*}
  \left\langle
    \varphi_{\sN}'^{2}(n_{\mu})
    \varphi_{\sN}'(n_{\mu}')
    \varphi_{\sN}'(n_{\mu}'')
  \right\rangle_{0}
  & = &
  \sigma_{\sN}^{2}\,
  g_{\sN}(n_{\mu}'-n_{\mu}'')
  +
  2\,
  g_{\sN}(n_{\mu}-n_{\mu}')\,
  g_{\sN}(n_{\mu}-n_{\mu}''),
  \\
  \left\langle
    \varphi_{\sN}'^{4}(n_{\mu})
    \varphi_{\sN}'(n_{\mu}')
    \varphi_{\sN}'(n_{\mu}'')
  \right\rangle_{0}
  & = &
  3\sigma_{\sN}^{4}\,
  g_{\sN}(n_{\mu}'-n_{\mu}'')
  +
  12\sigma_{\sN}^{2}\,
  g_{\sN}(n_{\mu}-n_{\mu}')\,
  g_{\sN}(n_{\mu}-n_{\mu}''),
\end{eqnarray*}

\noindent
as one can also see in Appendix~\ref{TabofInts},
Equations~(\ref{DecompFN2FN1FN1}) and~(\ref{DecompFN4FN1FN1}), in order to
write for our expectation value

\noindent
\begin{eqnarray*}
  \left\langle
    \varphi_{\sN}'(n_{\mu}')
    \varphi_{\sN}'(n_{\mu}'')
    S_{V}[\vec{\varphi}']
  \right\rangle_{0}
  & = &
  \sum_{n_{\mu}}^{N^{d}}
  \left\{
    \rule{0em}{5ex}
    \frac{\alpha-\alpha_{0}+\lambda v_{0}^{2}}{2}\,
    (\sN-1)
    \sigma_{0}^{2}\,
    g_{\sN}(n_{\mu}'-n_{\mu}'')
    +
  \right.
  \\
  &   &
  \hspace{2.42em}
  \left.
    +
    \frac{\alpha-\alpha_{\sN}+3\lambda v_{0}^{2}}{2}\,
    \sigma_{\sN}^{2}\,
    g_{\sN}(n_{\mu}'-n_{\mu}'')
    +
  \right.
  \\
  &   &
  \hspace{2.2em}
  \left.
    \rule{0em}{5ex}
    +
    \left[
      \alpha-\alpha_{\sN}+3\lambda v_{0}^{2}
    \right]
    g_{\sN}(n_{\mu}-n_{\mu}')\,
    g_{\sN}(n_{\mu}-n_{\mu}'')
    +
  \right.
  \\
  &   &
  \hspace{2.2em}
  \left.
    \rule{0em}{5ex}
    +
    \frac{\lambda}{4}\,
    (\sN^{2}-1)
    \sigma_{0}^{4}\,
    g_{\sN}(n_{\mu}'-n_{\mu}'')
    +
  \right.
  \\
  &   &
  \hspace{2.2em}
  \left.
    \rule{0em}{5ex}
    +
    \frac{\lambda}{2}\,
    (\sN-1)
    \sigma_{0}^{2}
    \sigma_{\sN}^{2}\,
    g_{\sN}(n_{\mu}'-n_{\mu}'')
    +
  \right.
  \\
  &   &
  \hspace{2.2em}
  \left.
    \rule{0em}{5ex}
    +
    \lambda
    (\sN-1)
    \sigma_{0}^{2}\,
    g_{\sN}(n_{\mu}-n_{\mu}')\,
    g_{\sN}(n_{\mu}-n_{\mu}'')
    +
  \right.
  \\
  &   &
  \hspace{2.2em}
  \left.
    \rule{0em}{5ex}
    +
    \frac{3\lambda}{4}\,
    \sigma_{\sN}^{4}\,
    g_{\sN}(n_{\mu}'-n_{\mu}'')
    +
  \right.
  \\
  &   &
  \hspace{2.2em}
  \left.
    \rule{0em}{5ex}
    +
    3\lambda
    \sigma_{\sN}^{2}\,
    g_{\sN}(n_{\mu}-n_{\mu}')\,
    g_{\sN}(n_{\mu}-n_{\mu}'')
  \right\}.
\end{eqnarray*}

\noindent
Next we group all terms containing $g_{\sN}(n_{\mu}'-n_{\mu}'')$ and
simplify to get

\noindent
\begin{eqnarray*}
  \lefteqn
  {
    \left\langle
      \varphi_{\sN}'(n_{\mu}')
      \varphi_{\sN}'(n_{\mu}'')
      S_{V}[\vec{\varphi}']
    \right\rangle_{0}
  }
  &   &
  \\
  & = &
  \sum_{n_{\mu}}^{N^{d}}
  \left[
    \rule{0em}{4ex}
    \frac{\alpha-\alpha_{0}+\lambda v_{0}^{2}}{2}\,
    (\sN-1)
    \sigma_{0}^{2}
    +
    \frac{\alpha-\alpha_{\sN}+3\lambda v_{0}^{2}}{2}\,
    \sigma_{\sN}^{2}
    +
  \right.
  \\
  &   &
  \hspace{2.2em}
  \left.
    \rule{0em}{4ex}
    +
    \frac{\lambda}{4}\,
    (\sN^{2}-1)
    \sigma_{0}^{4}
    +
    \frac{\lambda}{2}\,
    (\sN-1)
    \sigma_{0}^{2}
    \sigma_{\sN}^{2}
    +
    \frac{3\lambda}{4}\,
    \sigma_{\sN}^{4}
  \right]
  g_{\sN}(n_{\mu}'-n_{\mu}'')
  \\
  &   &
  +
  \sum_{n_{\mu}}^{N^{d}}
  \left\{
    \rule{0em}{3ex}
    \left[
      \alpha-\alpha_{\sN}+3\lambda v_{0}^{2}
    \right]
    +
    \lambda
    (\sN-1)
    \sigma_{0}^{2}
    +
    3\lambda
    \sigma_{\sN}^{2}
  \right\}
  g_{\sN}(n_{\mu}-n_{\mu}')\,
  g_{\sN}(n_{\mu}-n_{\mu}'').
\end{eqnarray*}

\noindent
The sum over $n_{\mu}$ can now be done in all terms of the first group,
yielding

\noindent
\begin{eqnarray*}
  \lefteqn
  {
    \left\langle
      \varphi_{\sN}'(n_{\mu}')
      \varphi_{\sN}'(n_{\mu}'')
      S_{V}[\vec{\varphi}']
    \right\rangle_{0}
  }
  &   &
  \\
  & = &
  N^{d}
  \left[
    \frac{\alpha-\alpha_{0}+\lambda v_{0}^{2}}{2}\,
    (\sN-1)
    \sigma_{0}^{2}
    +
    \frac{\alpha-\alpha_{\sN}+3\lambda v_{0}^{2}}{2}\,
    \sigma_{\sN}^{2}
    +
  \right.
  \\
  &   &
  \hspace{2.2em}
  \left.
    +
    \frac{\lambda}{4}\,
    (\sN^{2}-1)
    \sigma_{0}^{4}
    +
    \frac{\lambda}{2}\,
    (\sN-1)
    \sigma_{0}^{2}
    \sigma_{\sN}^{2}
    +
    \frac{3\lambda}{4}\,
    \sigma_{\sN}^{4}
  \right]
  g_{\sN}(n_{\mu}'-n_{\mu}'')
  \\
  &   &
  +
  \left[
    \alpha-\alpha_{\sN}
    +
    3\lambda
    v_{0}^{2}
    +
    \lambda
    (\sN-1)
    \sigma_{0}^{2}
    +
    3\lambda
    \sigma_{\sN}^{2}
  \right]
  \sum_{n_{\mu}}^{N^{d}}
  g_{\sN}(n_{\mu}-n_{\mu}')\,
  g_{\sN}(n_{\mu}-n_{\mu}'').
\end{eqnarray*}

\noindent
We must now perform the sum indicated. We get from
Appendix~\ref{TabofInts}, Equation~(\ref{DualPropNX}),

\begin{displaymath}
  \sum_{n_{\mu}}^{N^{d}}
  g_{\sN}(n_{\mu}-n_{\mu}')\,
  g_{\sN}(n_{\mu}-n_{\mu}'')
  =
  \frac{1}{N^{d}}
  \sum_{k_{\mu}}^{N^{d}}
  \frac
  {
    e^{-\ii(2\pi/N)\sum_{\mu}^{d}k_{\mu}(n_{\mu}'-n_{\mu}'')}
  }
  {
    [\rho^{2}(k_{\mu})+\alpha_{\sN}]^{2}
  }.
\end{displaymath}

\noindent
We have therefore the final result

\noindent
\begin{eqnarray}\label{ExpecValFNFNSV}
  \lefteqn
  {
    \left\langle
      \varphi_{\sN}'(n_{\mu}')
      \varphi_{\sN}'(n_{\mu}'')
      S_{V}[\vec{\varphi}']
    \right\rangle_{0}
  }
  &   &
  \nonumber\\
  & = &
  N^{d}
  \left[
    \frac{\alpha-\alpha_{0}+\lambda v_{0}^{2}}{2}\,
    (\sN-1)
    \sigma_{0}^{2}
    +
    \frac{\alpha-\alpha_{\sN}+3\lambda v_{0}^{2}}{2}\,
    \sigma_{\sN}^{2}
    +
  \right.
  \nonumber\\
  &   &
  \hspace{2.2em}
  \left.
    +
    \frac{\lambda}{4}\,
    (\sN^{2}-1)
    \sigma_{0}^{4}
    +
    \frac{\lambda}{2}\,
    (\sN-1)
    \sigma_{0}^{2}
    \sigma_{\sN}^{2}
    +
    \frac{3\lambda}{4}\,
    \sigma_{\sN}^{4}
  \right]
  g_{\sN}(n_{\mu}'-n_{\mu}'')
  \\
  &   &
  +
  \left[
    \alpha-\alpha_{\sN}
    +
    3\lambda
    v_{0}^{2}
    +
    \lambda
    (\sN-1)
    \sigma_{0}^{2}
    +
    3\lambda
    \sigma_{\sN}^{2}
  \right]
  \frac{1}{N^{d}}
  \sum_{k_{\mu}}^{N^{d}}
  \frac
  {
    e^{-\ii(2\pi/N)\sum_{\mu}^{d}k_{\mu}(n_{\mu}'-n_{\mu}'')}
  }
  {
    [\rho^{2}(k_{\mu})+\alpha_{\sN}]^{2}
  }.
  \nonumber
\end{eqnarray}

\section{Table of Integrals and Lattice Sums}\label{TabofInts}

\setcounter{equation}{0}

We give here a series of formulas and derivations involving Gaussian
integrals, Gaussian expectation values and lattice sums, in the context
the model discussed in this paper, which are used for the calculations
presented. All these can be derived from the basic result in momentum
space

\begin{equation}\label{BasicPropK}
  \left\langle
    \tvphi_{i}'(k_{\mu})
    \tvphi_{i}'^{*}(k_{\mu})
  \right\rangle_{0}
  =
  \frac{1}{N^{d}}\,
  \frac{1}{\rho^{2}(k_{\mu})+\alpha_{i}},
\end{equation}

\noindent
where $\alpha_{i}$ is either $\alpha_{0}$ or $\alpha_{\sN}$, depending on
the field component involved, and where $\rho^{2}(k_{\mu})$ are the
eigenvalues of the discrete Laplacian on the lattice, which are given by

\begin{displaymath}
  \rho^{2}(k_{\mu})
  =
  4
  \left[
    \sin^{2}\!\left(\frac{\pi k_{1}}{N}\right)
    +
    \ldots
    +
    \sin^{2}\!\left(\frac{\pi k_{d}}{N}\right)
  \right].
\end{displaymath}

\noindent
Since in the measure of $S_{0}[\vec{\varphi}']$ the modes are decoupled in
momentum space, the same expectation value with two different momenta
$k_{\mu}$ and $k_{\mu}'$ is zero by simple parity arguments. We use the
notation for the two-point functions in position space,

\noindent
\begin{eqnarray*}
  g_{0}(n_{\mu}-n_{\mu}')
  & = &
  \left\langle
    \varphi_{i}'(n_{\mu})\varphi_{i}'(n_{\mu}')
  \right\rangle_{0},
  \\
  g_{\sN}(n_{\mu}-n_{\mu}')
  & = &
  \left\langle
    \varphi_{\sN}'(n_{\mu})\varphi_{\sN}'(n_{\mu}')
  \right\rangle_{0},
\end{eqnarray*}

\noindent
for $i=1,\ldots,\sN-1$. These are, of course, the inverse Fourier
transforms of the corresponding two-point functions in momentum space,

\noindent
\begin{eqnarray*}
  g_{0}(n_{\mu}-n_{\mu}')
  & = &
  \sum_{k_{\mu}}^{N^{d}}
  e^{-\ii(2\pi/N)\sum_{\mu}^{d}k_{\mu}(n_{\mu}-n_{\mu}')}
  \left\langle
    \tvphi_{i}'(k_{\mu})
    \tvphi_{i}'^{*}(k_{\mu})
  \right\rangle_{0},
  \\
  g_{\sN}(n_{\mu}-n_{\mu}')
  & = &
  \sum_{k_{\mu}}^{N^{d}}
  e^{-\ii(2\pi/N)\sum_{\mu}^{d}k_{\mu}(n_{\mu}-n_{\mu}')}
  \left\langle
    \tvphi_{\sN}'(k_{\mu})
    \tvphi_{\sN}'^{*}(k_{\mu})
  \right\rangle_{0}.
\end{eqnarray*}

\noindent
In order to write this explicitly we may use the Fourier transforms of the
fields, for example in the case of the $\varphi_{\sN}'(n_{\mu})$ field
component,

\noindent
\begin{eqnarray*}
  g_{\sN}(n_{\mu}-n_{\mu}')
  & = &
  \left\langle
    \varphi_{\sN}'(n_{\mu})\varphi_{\sN}'(n_{\mu}')
  \right\rangle_{0}
  \\
  & = &
  \sum_{k_{\mu}}^{N^{d}}
  \sum_{k_{\mu}'}^{N^{d}}
  e^{-\ii(2\pi/N)\sum_{\mu}^{d}(k_{\mu}n_{\mu}+k_{\mu}'n_{\mu}')}
  \left\langle
    \tvphi_{\sN}'(k_{\mu})
    \tvphi_{\sN}'(k'_{\mu})
  \right\rangle_{0}.
\end{eqnarray*}

\noindent
The expectation value in momentum space in non-zero only if we have
$k_{\mu}'=-k_{\mu}$, in which case we have the result, which can be
obtained from Equation~(\ref{BasicPropK}) above,

\noindent
\begin{eqnarray*}
  \left\langle
    \tvphi_{\sN}'(k_{\mu})
    \tvphi_{\sN}'(-k_{\mu})
  \right\rangle_{0}
  & = &
  \left\langle
    \tvphi_{\sN}'(k_{\mu})
    \tvphi_{\sN}'^{*}(k_{\mu})
  \right\rangle_{0}
  \\
  & = &
  \frac{1}{N^{d}}\,
  \frac{1}{\rho^{2}(k_{\mu})+\alpha_{\sN}}.
\end{eqnarray*}

\noindent
This eliminates one of the momentum-space sums, and thus we get

\begin{equation}\label{SinglePropNX}
  g_{\sN}(n_{\mu}-n_{\mu}')
  =
  \frac{1}{N^{d}}
  \sum_{k_{\mu}}^{N^{d}}
  \frac
  {e^{-\ii(2\pi/N)\sum_{\mu}^{d}k_{\mu}(n_{\mu}-n_{\mu}')}}
  {\rho^{2}(k_{\mu})+\alpha_{\sN}},
\end{equation}

\noindent
which is just the statement that $g_{\sN}(n_{\mu}-n_{\mu}')$ is the
inverse Fourier transform of the momentum-space propagator. Note that this
is necessarily real, and that therefore the imaginary part of the
right-hand side vanishes. In a completely similar way, we have the
corresponding result for the other field $\sN-1$ components, with
$i\neq\sN$,

\begin{displaymath}
  g_{0}(n_{\mu}-n_{\mu}')
  =
  \frac{1}{N^{d}}
  \sum_{k_{\mu}}^{N^{d}}
  \frac
  {e^{-\ii(2\pi/N)\sum_{\mu}^{d}k_{\mu}(n_{\mu}-n_{\mu}')}}
  {\rho^{2}(k_{\mu})+\alpha_{0}}.
\end{displaymath}

\noindent
The following sum involving $g_{0}(n_{\mu}-n_{\mu}')$ can also be easily
calculated, using once more the Fourier transforms,

\begin{displaymath}
  \sum_{n_{\mu}}^{N^{d}}g_{\sN}(n_{\mu}-n_{\mu}')
  =
  \frac{1}{N^{d}}
  \sum_{n_{\mu}}^{N^{d}}
  \sum_{k_{\mu}}^{N^{d}}
  \frac
  {e^{-\ii(2\pi/N)\sum_{\mu}^{d}k_{\mu}(n_{\mu}'-n_{\mu}'')}}
  {\rho^{2}(k_{\mu})+\alpha_{\sN}}.
\end{displaymath}

\noindent
The orthogonality relation can be used to simplify this expression, and
thus we get

\noindent
\begin{eqnarray*}
  \sum_{n_{\mu}}^{N^{d}}g_{\sN}(n_{\mu}-n_{\mu}')
  & = &
  \sum_{k_{\mu}}^{N^{d}}
  \delta^{d}(k_{\mu},0_{\mu})\,
  \frac
  {e^{\ii(2\pi/N)\sum_{\mu}^{d}k_{\mu}n_{\mu}'}}
  {\rho^{2}(k_{\mu})+\alpha_{\sN}}
  \\
  & = &
  \frac{e^{0}}{\rho^{2}(0)+\alpha_{\sN}}
  \\
  & = &
  \frac{1}{\alpha_{\sN}}.
\end{eqnarray*}

\noindent
This is simply the zero-mode of the propagator. The same can be done for
the other components of the field, so we conclude that

\noindent
\begin{eqnarray}
  \sum_{n_{\mu}}^{N^{d}}
  g_{0}(n_{\mu}-n_{\mu}')
  =
  \frac{1}{\alpha_{0}},
  \nonumber\\
  \label{SumgN}
  \sum_{n_{\mu}}^{N^{d}}
  g_{\sN}(n_{\mu}-n_{\mu}')
  =
  \frac{1}{\alpha_{\sN}}.
\end{eqnarray}

\noindent
A similar sum with two chained factors of $g_{\sN}(n_{\mu}-n_{\mu}')$ can
be calculated in a similar way. Using the Fourier expressions of
$g_{0}(n_{\mu}-n_{\mu}')$ and $g_{0}(n_{\mu}-n_{\mu}'')$ we get

\noindent
\begin{eqnarray*}
  \lefteqn
  {
    \sum_{n_{\mu}}^{N^{d}}
    g_{0}(n_{\mu}-n_{\mu}')\,
    g_{0}(n_{\mu}-n_{\mu}'')
  }
  &   &
  \\
  & = &  
  \sum_{n_{\mu}}^{N^{d}}
  \frac{1}{N^{2d}}
  \sum_{k_{\mu}'}^{N^{d}}
  \sum_{k_{\mu}''}^{N^{d}}
  \frac
  {
    e^{-\ii(2\pi/N)\sum_{\mu}^{d}[k_{\mu}'(n_{\mu}-n_{\mu}')+k_{\mu}''(n_{\mu}-n_{\mu}'']}
  }
  {
    [\rho^{2}(k_{\mu}')+\alpha_{0}]
    [\rho^{2}(k_{\mu}'')+\alpha_{0}]
  }
  \\
  & = &  
  \frac{1}{N^{2d}}
  \sum_{k_{\mu}'}^{N^{d}}
  \sum_{k_{\mu}''}^{N^{d}}
  \frac
  {
    e^{\ii(2\pi/N)\sum_{\mu}^{d}(k_{\mu}'n_{\mu}'+k_{\mu}''n_{\mu}'')}
  }
  {
    [\rho^{2}(k_{\mu}')+\alpha_{0}][\rho^{2}(k_{\mu}'')+\alpha_{0}]
  }
  \sum_{n_{\mu}}^{N^{d}}
  e^{-\ii(2\pi/N)\sum_{\mu}^{d}(k_{\mu}'+k_{\mu}'')n_{\mu}}
  \\
  & = &  
  \frac{1}{N^{d}}
  \sum_{k_{\mu}'}^{N^{d}}
  \sum_{k_{\mu}''}^{N^{d}}
  \delta^{d}(k_{\mu}',-k_{\mu}'')\,
  \frac
  {
    e^{\ii(2\pi/N)\sum_{\mu}^{d}(k_{\mu}'n_{\mu}'+k_{\mu}''n_{\mu}'')}
  }
  {
    [\rho^{2}(k_{\mu}')+\alpha_{0}][\rho^{2}(k_{\mu}'')+\alpha_{0}]
  }
  \\
  & = &  
  \frac{1}{N^{d}}
  \sum_{k_{\mu}''}^{N^{d}}
  \frac
  {
    e^{-\ii(2\pi/N)\sum_{\mu}^{d}(k_{\mu}''n_{\mu}'-k_{\mu}''n_{\mu}'')}
  }
  {
    [\rho^{2}(k_{\mu}'')+\alpha_{0}]^{2}
  }.
\end{eqnarray*}

\noindent
We see therefore that we get the sum expressed as a Fourier transform,
with the general structure of a two-point function,

\begin{equation}\label{DualProp0X}
  \sum_{n_{\mu}}^{N^{d}}
  g_{0}(n_{\mu}-n_{\mu}')\,
  g_{0}(n_{\mu}-n_{\mu}'')
  =
  \frac{1}{N^{d}}
  \sum_{k_{\mu}}^{N^{d}}
  \frac
  {
    e^{-\ii(2\pi/N)\sum_{\mu}^{d}k_{\mu}(n_{\mu}'-n_{\mu}'')}
  }
  {
    [\rho^{2}(k_{\mu})+\alpha_{0}]^{2}
  }.
\end{equation}

\noindent
A similar result is true, of course, for the remaining field component

\begin{equation}\label{DualPropNX}
  \sum_{n_{\mu}}^{N^{d}}
  g_{\sN}(n_{\mu}-n_{\mu}')\,
  g_{\sN}(n_{\mu}-n_{\mu}'')
  =
  \frac{1}{N^{d}}
  \sum_{k_{\mu}}^{N^{d}}
  \frac
  {
    e^{-\ii(2\pi/N)\sum_{\mu}^{d}k_{\mu}(n_{\mu}'-n_{\mu}'')}
  }
  {
    [\rho^{2}(k_{\mu})+\alpha_{\sN}]^{2}
  }.
\end{equation}

\noindent
The squared dispersions, also referred to as widths or variances of the
fields at a given site, are denoted as

\noindent
\begin{eqnarray*}
  \sigma_{0}^{2}
  & = &
  \left\langle
    \varphi_{i}'^{2}(n_{\mu})
  \right\rangle_{0},
  \\
  \sigma_{\sN}^{2}
  & = &
  \left\langle
    \varphi_{\sN}'^{2}(n_{\mu})
  \right\rangle_{0},
\end{eqnarray*}

\noindent
for $i=1,\ldots,\sN-1$. Using the expression of the two-point function in
terms of Fourier components we may write these explicitly as

\noindent
\begin{eqnarray}
  \label{Sigma0Form}
  \sigma_{0}^{2}
  & = &
  \frac{1}{N^{d}}
  \sum_{k_{\mu}}^{N^{d}}
  \frac{1}{\rho^{2}(k_{\mu})+\alpha_{0}},
  \\
  \label{SigmaNForm}
  \sigma_{\sN}^{2}
  & = &
  \frac{1}{N^{d}}
  \sum_{k_{\mu}}^{N^{d}}
  \frac{1}{\rho^{2}(k_{\mu})+\alpha_{\sN}}.
\end{eqnarray}

\noindent
In terms of these quantities the following decompositions of higher-point
functions can be established, always for $i=1,\ldots,\sN-1$,

\noindent
\begin{eqnarray}
  \left\langle
    \varphi_{i}'^{4}(n_{\mu})
  \right\rangle_{0}
  & = &
  3\sigma_{0}^{4},
  \nonumber\\
  \label{DecompFN4}
  \left\langle
    \varphi_{\sN}'^{4}(n_{\mu})
  \right\rangle_{0}
  & = &
  3\sigma_{\sN}^{4},
  \\
  \left\langle
    \varphi_{i}'^{3}(n_{\mu})
    \varphi_{i}'(n_{\mu}')
  \right\rangle_{0}
  & = &
  3\sigma_{0}^{2}\,
  g_{0}(n_{\mu}-n_{\mu}'),
  \nonumber\\
  \label{DecompFN3FN1}
  \left\langle
    \varphi_{\sN}'^{3}(n_{\mu})
    \varphi_{\sN}'(n_{\mu}')
  \right\rangle_{0}
  & = &
  3\sigma_{\sN}^{2}\,
  g_{\sN}(n_{\mu}-n_{\mu}'),
  \\
  \label{DecompFi2Fi1Fi1}
  \left\langle
    \varphi_{i}'^{2}(n_{\mu})
    \varphi_{i}'(n_{\mu}')
    \varphi_{i}'(n_{\mu}'')
  \right\rangle_{0}
  & = &
  \sigma_{0}^{2}\,
  g_{0}(n_{\mu}'-n_{\mu}'')
  +
  \nonumber\\
  &   &
  +
  2\,
  g_{0}(n_{\mu}-n_{\mu}')\,
  g_{0}(n_{\mu}-n_{\mu}''),
  \\
  \label{DecompFN2FN1FN1}
  \left\langle
    \varphi_{\sN}'^{2}(n_{\mu})
    \varphi_{\sN}'(n_{\mu}')
    \varphi_{\sN}'(n_{\mu}'')
  \right\rangle_{0}
  & = &
  \sigma_{\sN}^{2}\,
  g_{\sN}(n_{\mu}'-n_{\mu}'')
  +
  \nonumber\\
  &   &
  +
  2\,
  g_{\sN}(n_{\mu}-n_{\mu}')\,
  g_{\sN}(n_{\mu}-n_{\mu}''),
  \\
  \label{DecompFi4Fi1Fi1}
  \left\langle
    \varphi_{i}'^{4}(n_{\mu})
    \varphi_{i}'(n_{\mu}')
    \varphi_{i}'(n_{\mu}'')
  \right\rangle_{0}
  & = &
  3\sigma_{0}^{4}\,
  g_{0}(n_{\mu}'-n_{\mu}'')
  +
  \nonumber\\
  &   &
  +
  12\sigma_{0}^{2}\,
  g_{0}(n_{\mu}-n_{\mu}')\,
  g_{0}(n_{\mu}-n_{\mu}''),
  \\
  \label{DecompFN4FN1FN1}
  \left\langle
    \varphi_{\sN}'^{4}(n_{\mu})
    \varphi_{\sN}'(n_{\mu}')
    \varphi_{\sN}'(n_{\mu}'')
  \right\rangle_{0}
  & = &
  3\sigma_{\sN}^{4}\,
  g_{\sN}(n_{\mu}'-n_{\mu}'')
  +
  \nonumber\\
  &   &
  +
  12\sigma_{\sN}^{2}\,
  g_{\sN}(n_{\mu}-n_{\mu}')\,
  g_{\sN}(n_{\mu}-n_{\mu}'').
\end{eqnarray}

\noindent
It is also not difficult to expand and calculate the following sums,

\noindent
\begin{eqnarray*}
  \left\langle
    \left[
      \sum_{i=2}^{\sN-1}
      \varphi_{i}'^{2}(n_{\mu})
    \right]^{2}
  \right\rangle_{0}
  & = &
  (\sN-2)
  \left\langle
    \varphi_{1}'^{4}(n_{\mu})
  \right\rangle_{0}
  +
  (\sN-2)(\sN-3)
  \left\langle
    \varphi_{1}'^{2}(n_{\mu})
  \right\rangle_{0}^{2}
  \\
  & = &
  3(\sN-2)
  \sigma_{0}^{4}
  +
  (\sN-2)(\sN-3)
  \sigma_{0}^{4}
  \\
  & = &
  \rule{0em}{4ex}
  (\sN)(\sN-2)
  \sigma_{0}^{4},
  \\
  \left\langle
    \left[
      \sum_{i=1}^{\sN-1}
      \varphi_{i}'^{2}(n_{\mu})
    \right]^{2}
  \right\rangle_{0}
  & = &
  (\sN-1)
  \left\langle
    \varphi_{1}'^{4}(n_{\mu})
  \right\rangle_{0}
  +
  (\sN-1)(\sN-2)
  \left\langle
    \varphi_{1}'^{2}(n_{\mu})
  \right\rangle_{0}^{2}
  \\
  & = &
  3(\sN-1)
  \sigma_{0}^{4}
  +
  (\sN-1)(\sN-2)
  \sigma_{0}^{4}
  \\
  & = &
  \rule{0em}{4ex}
  (\sN+1)(\sN-1)
  \sigma_{0}^{4},
\end{eqnarray*}

\noindent
so that we get the results

\noindent
\begin{eqnarray}
  \label{SumFiN-2}
  \left\langle
    \left[
      \sum_{i=2}^{\sN-1}
      \varphi_{i}'^{2}(n_{\mu})
    \right]^{2}
  \right\rangle_{0}
  & = &
  \sN(\sN-2)
  \sigma_{0}^{4},
  \\
  \label{SumFiN-1}
  \left\langle
    \left[
      \sum_{i=1}^{\sN-1}
      \varphi_{i}'^{2}(n_{\mu})
    \right]^{2}
  \right\rangle_{0}
  & = &
  (\sN+1)(\sN-1)
  \sigma_{0}^{4}.
\end{eqnarray}

\end{document}